\documentclass[10pt,aps,pra,twocolumn,superscriptaddress,floatfix,nofootinbib]{revtex4-2}

\usepackage{amsmath, amsfonts, amssymb}

\usepackage{bm,bbm, braket, accents}
\usepackage{graphicx}   % need for figures
\usepackage[usenames,dvipsnames]{color}
\usepackage{lipsum} 
\usepackage{tikz}
\usetikzlibrary{quantikz}
\usepackage{pgfplots}
\pgfplotsset{compat=newest}
\usepackage{pifont}

\makeatletter
\def\@bibdataout@aps{%
 \immediate\write\@bibdataout{%
  @CONTROL{%
   apsrev41Control,author="08",editor="1",pages="0",title="0",year="1",eprint="1"%
  }%
 }%
 \if@filesw
  \immediate\write\@auxout{\string\citation{apsrev41Control}}%
 \fi
}%
\makeatother % Phew.
\newcommand{\dg}{^\dagger}

\newcommand{\ip}[2]{\langle{#1}|{#2}\rangle}
\newcommand{\op}[2]{\ket{#1}\!\bra{#2}}

\newcommand{\ZN}{\hat{Z}_N}
\newcommand{\XN}{\hat{X}_N}
\newcommand{\SN}{\hat{S}_N}
\newcommand{\TN}{\hat{T}_N}

\newcommand{\zeroL}{0_N}
\newcommand{\oneL}{1_N}
\newcommand{\plusL}{+_N}
\newcommand{\minusL}{-_N}
\newcommand{\CROT}{{\rm CROT}}
\newcommand{\CCROT}{{\rm CCROT}}

\newcommand{\CPHASE}{\textsc{cphase}}

\newcommand{\E}[2]{\hat{\mathbb{E}}_{#1}\!\left( #2 \right)}

\newcommand{\lbar}{\bar}

\newcommand{\X}{\lbar X}

\newcommand{\Z}{\lbar Z}

\newcommand{\CZ}{\overline{\rm CZ}}
\newcommand{\CCZ}{\overline{\rm CCZ}}

\newcommand{\Sgate}{\lbar S}

\newcommand{\phaseOP}{\hat{\Sigma}}

\usepackage[breaklinks=true]{hyperref}
\hypersetup{
  colorlinks   = true, %Colour links instead of boxes
  urlcolor     = blue, %Colour for external hyperlinks
  linkcolor    = blue, %Colour of internal links
  citecolor    = red %Colour of citations
}

\usepackage[normalem]{ulem}

\usepackage[capitalise]{cleveref} % must be loaded after hyperref
\crefformat{equation}{Eq.~(#2#1#3)} % These change 'equation' to Eq., more PRL or PRA-style
\crefformat{section}{Sec.~#2#1#3} % These change 'equation' to Eq., morePRL or PRA-style
\Crefformat{equation}{Equation~(#2#1#3)}
\crefformat{figure}{Fig.~#2#1#3}
\crefrangeformat{equation}{Eqs.~#3(#1)#4--#5(#2)#6}
\Crefformat{section}{Section~#2#1#3}

\usepackage{graphicx}

\begin{document}
%============================================

\title{An explicit error correction scheme and code distance for bosonic codes with rotational symmetry}

\author{Benjamin Marinoff}
\affiliation{Department of Physics, University of Colorado, Boulder, Colorado 80309, USA}

\author{Miles Bush}
\affiliation{Department of Electrical, Computer and Energy Engineering, University of Colorado, Boulder, Colorado 80309, USA}

\author{Joshua Combes}
%\email{joshua.combes@gmail.com}
\affiliation{Department of Electrical, Computer and Energy Engineering, University of Colorado, Boulder, Colorado 80309, USA}

\date{\today}

%============================================

\begin{abstract}
Bosonic codes with rotational symmetry are currently one of the best performing quantum error correcting codes. Little is known about error propagation and code distance for these rotation codes in contrast with qubit codes and Bosonic codes with translation symmetry. We use a general purpose error basis that is naturally suited to codes with rotation symmetry to compute how errors propagate through gates. This error basis allows us to give an explicit error detection, decoding, and correction scheme for any code with rotation symmetry. We also prove that codes with an $N$-fold rotation symmetry have a distance of $(d_n, d_\theta)=(N, \pi/N)$ with respect to number and rotation errors. 
\end{abstract}

%============================================

\maketitle

%============================================
\section{Introduction} \label{sec:intro}
%============================================

Experimentally, codes with rotation symmetry~\cite{GrimCombBara20,xu_clifford_2023} have come close to the break even point~\cite{Ofek,LyanSun} where logical error rates are no worse than the physical error rates on the same hardware when manipulating unencoded information. Recent evidence shows that codes with translation symmetry~\cite{GKP2001,tosta_grand_2022} have also achieved~\cite{Campagne-Ibarcq2020} and surpassed~\cite{Sivak2023} this important milestone on the path towards quantum fault tolerance.

Codes with translation symmetry, have many nice properties that make them convenient to work with. In particular, many logical operations and stabilizers can be understood through the lens of displacement operators~\cite{GKP2001}. Moreover, these displacement operators form a basis for the Hilbert space of a quantum harmonic oscillator~\cite{CahillGlauber69}, so any arbitrary error can be decomposed into a linear combination of such displacements. This gives an intuitive picture of error detection and then correction based on displacement operators.

The goal of this work is to develop a similarly simple description for rotation codes. Specifically, we will define code distance and develop an intuitive error detection and correction scheme for rotation codes, utilizing an appropriately chosen error basis. We begin by summarizing the mathematics of bosonic codes with rotation symmetry in \cref{sec:summary} by defining the code space, logical gates, stabilizers, and an error basis. The error basis, which contains products of phase space rotations and numerical shifts in photon number, is well suited for rotation codes just as displacements are well suited for translation codes. Many of the new bosonic gates introduced in this section may be implemented via a SNAP gate \cite{SNAPgateTHY2015,SNAPgateEXP2015}. In \cref{sec:errprop} we examine how errors propagate through various logical gates, rederiving the results from \citet{GrimCombBara20} in the new error basis and extending them to a larger collection of gates. We consider an error that has occurred before the gate is applied and construct an equivalent circuit where the gate is applied to an uncorrupted state and then a different error occurs afterwards, as if the original error was ``commuted through'' the gate. In doing so, we can see if the gates compound errors in a way that prevents error correction. These calculations help inform which gate sets are nice for fault tolerant schemes, e.g. which gates preserve noise bias. Next, in \cref{sec:explicit_QEC} we use the above results to develop an explicit error correction scheme for rotational codes, including a naive decoder, which is as intuitive as that of translation codes. Finally, in \cref{sec:num_phase_trade} we use the new error basis to explicitly compute the distance of rotation codes and explore the trade off between resilience to shift errors and phase errors. This builds on the prior work of several authors~\cite{Michael16, albert2018, Yingkai2021}. The important caveat for \cref{sec:explicit_QEC,sec:num_phase_trade} is these results strictly hold only in the limit of infinite energy codes (which is analogous to the situation with translation codes). A brief conclusion follows in \cref{sec:conclusion}.

%============================================
\section{Review of rotation codes} \label{sec:summary}
%============================================
In this section we summarize the fundamentals of codes with rotation symmetry. Readers looking for more details should consult Ref.~\cite{GrimCombBara20}.

\begin{figure}
    \includegraphics[width=\columnwidth]{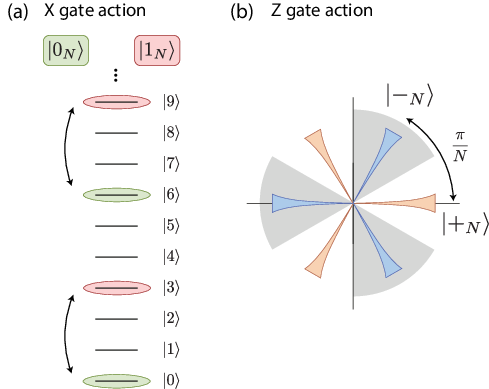}
    \caption{ Action of gates on rotation codes in Fock space and phase space. In (a), a truncated Fock grid is depicted for an $N=3$ code. The amplitudes of the code words $\ket{0_N}$ and $\ket{1_N}$ are depicted as green and red ellipses. A possible implementation of a logical $\X$ gate depicted by the black arrows. In (b)  we use a ball-and-stick like diagram to simplify the phase space (Wigner function) description of the states. %  $W(\alpha)$ The x axis is the real part of $\alpha$ and the y axis is the imaginary part.
The orange radial spikes correspond to the $\ket{+_N}$ state while $\ket{-_N}$ is in blue. The action of a $\ZN$ gate is a rotation by $\pi/N$.
}\label{fig:gcb_codes}
\end{figure}

A code has discrete $N$-fold rotation symmetry if any state $\ket\psi$ in the code subspace (codespace) is an eigenstate of the discrete rotation operator
\begin{equation}\label{eq:RN}
    \hat R_N = \exp \!\left [ i \frac{2\pi}{N}\hat n \right],
\end{equation}
where $\hat n = \hat a\dg \hat a$ is the Fock space number operator and $[\hat a,\hat a\dg]=1$. The square root of $\hat R_N$ is a Hermitian operator that squares to the identity on the codespace, which is invariant under $\hat{R}_N$. Thus the operator
\begin{equation}\label{eq:Z_N}
    \ZN \equiv \sqrt{\hat R_N} = \hat R_{2N} = \exp\! \left[ i \frac{\pi}{N}\hat n \right]
\end{equation}
preserves the rotation symmetry of $\ket{\psi}$. We define an {\em order-N bosonic rotation code} (henceforth called simply a rotation code), to be a code where the operator $\ZN$
acts as logical $\Z$ (and the choice to define this operator as $\Z$ rather than $\X$ is just a convention). In this convention the $\Z$ basis code words are of the form
\begin{subequations}\label{eq:logicalz_codewords}
\begin{align}
    \ket{0_N} ={}& \sum_{k=0}^\infty f_{2kN} \ket{2kN}, \label{eq:logicalz_0} \\
    \ket{1_N} ={}& \sum_{k=0}^\infty f_{(2k+1)N} \ket{(2k+1)N} \label{eq:logicalz_1} \,
\end{align}
\end{subequations}
here the kets $\ket{2kN}$ and $\ket{(2k+1)N}$ are eigenstates of the number operator $\hat{n}$, see \cref{fig:gcb_codes} (a).
The amplitudes $f_{2kN}$ and $f_{(2k+1)N}$ are the only thing that distinguish rotation codes from one another~\cite{GrimCombBara20} and differ between cat codes~\cite{Cochrane99,Ralph2003,Zaki,Mirrahimi_2014} and e.g. binomial codes~\cite{Michael16}. The dual-basis codewords, $\ket{\pm_N}$, are constructed as usual via superpositions of the computational basis codewords, $\ket{\pm_N} = \frac{1}{\sqrt{2}}(\ket{0_N} \pm \ket{1_N})$, yielding Fock space representations
\begin{subequations} \label{eq:logicalx}
\begin{align} 
    \ket{+_N} = & \frac{1}{\sqrt{2}}\sum_{k=0}^\infty  f_{kN} \ket{kN}\label{eq:logicalxplus},\\
    \ket{-_N} = & \frac{1}{\sqrt{2}} \sum_{k=0}^\infty  (-1)^k f_{kN} \ket{kN}\label{eq:logicalxminus}.
\end{align}
\end{subequations}
Both $\ket{\pm_N}$ have support on the full set of $\ket{kN}$ Fock states. It is evident that $\ZN$ acts as logical $\Z$ on these states:
\begin{equation}
    \ZN \ket{\pm_N} = \ket{\mp_N}\, ,
\end{equation}
this action is depicted in \cref{fig:gcb_codes} (b).

To find an approximate logical $\X$ gate we note that number and phase operators form an approximate conjugate pair, and since $\ZN$ is a complex exponential of $\hat{n}$, the complex exponential of a phase operator is likely a good candidate for $\X$. Such a phase operator has presented many historic obstacles which will not be detailed here; the curious reader is pointed towards Ref.~\cite{CarrNietRMP68}. 

In this paper we use the \citet{Suss64} phase operator. The Susskind and Glogower phase operator is easily expressed in the number basis as
\begin{subequations}
\begin{align}
    \phaseOP_1^- &= \widehat \exp(i\phi) =  \sum_{n=0}^{\infty} \op{n}{n+1}\\ \phaseOP_1^+ &= \widehat \exp(-i\phi)  = \big (\phaseOP_1^-\big )^\dagger
\end{align}
\end{subequations}
The hat over the exponential, rather than the argument of the exponential, is a standard notation used to remind the reader that these operators are {\bf not} exponentials of a Hermitian phase operator~\cite{CarrNietRMP68}. These operators act like the raising and lowering operators without the numeric $\sqrt{n}$  coefficients, i.e. $\phaseOP_1^-\ket{n} = \ket{n-1}$ and $\phaseOP_1^+\ket{n} = \ket{n+1}$, making them much easier to work with algebraically.
Moreover, if we examine the action of these operators on the (non-normalizable) London phase state \cite{CarrNietRMP68}
\begin{equation}\label{eq:london_phase_state}
    \ket{\varphi} = \sum_{n=0}^\infty e^{i\varphi n} \ket{n}
\end{equation}
we find that 
\begin{equation}
    \phaseOP_1^- \ket{\varphi} = \widehat \exp(i\phi) \ket{\varphi} = e^{i\varphi}\ket{\varphi}.    
\end{equation}

Because our codes have an $N$-fold rotational symmetry, the relevant phase variable can be defined on the interval $\phi \in [-\pi/N, \pi/N)$~\cite{CombesAlbertetal2021}. We call this variable the modular phase, and the associated modular phase operators are
\begin{subequations}\label{eq:pow_susskind}
\begin{align}
    \phaseOP_N^- &=  \big (\phaseOP_1^-\big)^N = \widehat \exp(iN\phi)= \sum_{n=0}^{\infty} \op{n}{n+N},\\
    \phaseOP_N^+ &= \big (\phaseOP_N^-\big)^\dagger = \widehat \exp(-iN\phi)= \sum_{n=0}^{\infty} \op{n+N}{n},
\end{align}
\end{subequations}
as described in detail in Ref.~\cite{CombesAlbertetal2021}. Following \cite{GrimCombBara20} we define the dual basis codewords of {\em ideal phase codes} to be rotated superpositions of London phase states
\begin{equation} \label{eq:ideal_code}
    \ket{\pm_N}_{\rm ideal} = \sum_{m=0}^{N-1}  (\pm 1)^m \left |\varphi = \frac{2m\pi}{N}\right \rangle.
\end{equation}
It turns out the dual basis codewords are best understood in phase space. In \cref{fig:gcb_codes}(b) we depict finite energy versions of these states in phase space. The non-normalizable states in \cref{eq:ideal_code} are eigenstates of the modular phase operators in ~\cref{eq:pow_susskind}. Since the underlying unnormalized states $\ket{\varphi}$ in \cref{eq:london_phase_state} have the property $|\ip{m}{\varphi}|^2 = 1$, we see that $\ip{mN}{\pm_N}_{\rm ideal} = f_{mN} = \text{const.}$ for all $m$. For such codes, $\phaseOP_N^-\ket{\pm_N}_{\rm ideal}= \pm \ket{\pm_N}_{\rm ideal}$, thus we can define
\begin{equation} \label{eq:X_gate}
    \XN = \phaseOP_{N}^- = \sum_{n=0}^{\infty} \op{n}{n+N}
\end{equation}
which acts as a logical $\X$ operator on the codespace as desired, and it can be seen that $\ZN$ and $\XN$ anticommute.

The commuting stabilizers on this codespace are 
\begin{subequations} \label{eq:stabilizers}
\begin{align}
    \hat S_{\Z} &= \hat R_N = e^{i (2\pi/N) \hat n} & {\rm (number)} \label{eq:stabilizerZ} \\
    \hat S_{\X} &= \phaseOP_{2N}^- = \sum_{n=0}^\infty \op{n}{n+2N} &{\rm (phase)}, \label{eq:stabilizerX}
\end{align}
\end{subequations}
which are called the number and phase stabilizers because of the errors their syndrome measurements diagnose, as seen in \cref{sec:explicit_QEC}. %Notice the phase stabilizer has double the rotation symmetry of $\ZN$ so a measurement of $\hat S_{\Z}$ can't distinguish between $\ket{\pm_N}$.

Finally, we define an {\em approximate phase code} to be a code whose variance in the operators in \cref{eq:pow_susskind} goes to zero in the limit of some parameter, thus the code reduces to an ideal phase code in that limit.

As an illustrative example, consider the family of cat codes~\cite{Cochrane99,Ralph2003,Zaki,Mirrahimi_2014} parametrized by coherent state $\ket{\alpha}$, defined by the codewords
\begin{align*}
    \ket{0_{N, \alpha}} &= \frac{1}{\sqrt{\mathcal N_0}} \sum_{m=0}^{2N-1} e^{i(m\pi /N)\hat n} \ket{\alpha} \\
    \ket{1_{N,\alpha}} &= 
    \frac{1}{\sqrt{\mathcal N_1}}
    \sum_{m=0}^{2N-1} (-1)^m e^{i(m\pi /N)\hat n} \ket{\alpha}.
\end{align*}
The cat codes are approximate phase codes because in the limit $|\alpha| \rightarrow \infty$, the variance in the modular phase operators goes to zero, as shown in Ref.~\cite{GrimCombBara20}. Thus in this limit the amplitudes $\ip{mN}{\pm_{N,\alpha}}$ become constant and the code is an ideal phase codes. Similarly, binomial codes also limit to ideal phase codes.

%============================================
\subsection{Computation and gate definitions}
%============================================
We now define a number of gates which are more than sufficient to enact a universal quantum computation scheme. For example a subset of the gates we examine were shown to be universal in Ref.~\cite{GrimCombBara20}, specifically the gates $\{\Sgate, \CZ\}$, preparations of two states $\{ \ket{+_N}, \ket{T_N}\}$ and a measurement in the logical $\ket{\pm_N}$ basis. 

In addition to $\ZN$ and $\XN$ (repeated here for convenience), we may define several other single-qubit gates:
\begin{subequations} \label{eq:singlequbitgates}
\begin{align}
    \ZN &= \hat R_{2N} =\exp\left [i \frac{\pi}{N} \hat{n} \right ] \label{eq:Z_gatey}\\
    \XN &= \hat \phaseOP_{N}^- = \sum_{n=0}^{\infty} \op{n}{n+N}\\
    \XN' &= \sum_{n=0}^{\infty}\sum_{\ell=0}^{N-1} \op{2nN+\ell}{(2n+1)N+\ell}+ {\rm h.c.} \label{eq:X_prime_gate}\\
    \SN &= \exp \left [i \frac{\pi}{2N^2} \hat{n}^2 \right ] \label{eq:S_gate} \\
    \TN &= \exp \left [i \frac{\pi}{4N^4} \hat{n}^4 \right ] \label{eq:T_gate} \\
    \TN' &= \exp \left [i \frac{\pi}{4}\left (2 \left(\frac{\hat{n}}{N}\right)^3 + \left(\frac{\hat{n}}{N}\right)^2 - 2 \frac{\hat{n}}{N}\right) \right ] \, ,\label{eq:T_gate_prime}
\end{align}
\end{subequations}
where h.c. means Hermitian conjugate. As discussed above, $\XN$ acts as an approximate $\X$ gate for approximate phase codes by shifting the number states down by $N$. To make this an exact $\X$ gate, however, would require re-weighting the amplitudes after the shift. This re-weighting is specific to the kind of rotation code used and can be quite complicated. An alternate implementation, $\XN'$, is proposed that shuffles the states reversibly rather than shifting them all down by $N$. This acts as an exact $\X$ gate on the codespace. We call this gate Baragiola's ``bin swap gate''~\cite{BaraG}. $\SN$ and $\TN$ are simple realizations of the $\pi/4$ and $\pi/8$ gates respectively. $\TN'$ is an alternate realization of the $\pi/8$ gate (introduced in Ref.~\cite{AlvCalcFerrFerr20}) that looks more complicated, but introduces smaller errors after propagation as it is lower order in $\hat{n}$.

We also consider the following multiqubit gates
\begin{subequations}
\begin{align}
    \CROT_{NM} & = \exp\left [i\frac{\pi}{NM} \hat n \otimes \hat n \right ] \\
    \CCROT_{NMO} & = \exp\left [i\frac{\pi}{NMO} \hat n \otimes \hat n \otimes \hat n \right ]
\end{align}
\end{subequations}
where for full generality we allow the different logical qubits to have distinct rotational symmetries $N,M,O$. The $\CROT$ gate acts as a logical $\CZ$ and $\CCROT$ acts as a logical $\CCZ$ on the code space. In some sense these operators can be thought of as conditional rotations by $\pi$, but an arbitrary conditional rotation of angle $\phi$ cannot be naively constructed by replacing $\pi$ with $\phi$. Constructions of gates that rotate about $\Z$ by an amount $\phi$ where $\phi\neq\pi$ are discussed in \cref{sec:rotation_gates}.

%============================================
\subsection{Error Basis}\label{sec:errbasis}
%============================================

An operator basis for qubit quantum computation is given by the $n$-fold tensor product of the Pauli operators $\{I,X,Y,Z \}^{\otimes n}$. For a bosonic Hilbert space it is known that the displacement operators form an operator basis~\cite{CahillGlauber69}. Here we use a newly introduced operator basis~\cite{CombesAlbertetal2021} that is naturally suited to bosonic codes with a rotation symmetry. 

Previously \citet{GrimCombBara20} introduced the operator basis 
\begin{equation}\label{eq:errormodel}
  \left\{e^{i\theta\hat n}\hat a^k,\, \left(\hat a\dg\right)^k e^{-i\theta \hat n} \right\}
\end{equation}
where $k\ge 0$ is an integer and $\theta \in[-\pi, \pi)$, which can be written as
\begin{equation}\label{eq:E}
  \hat E_k(\theta) \equiv \left\{ \begin{array}{ll}
      e^{i\theta\hat n}\hat a^{|k|} & \text{ for } k < 0 \\
      \left(\hat a\dg\right)\phantom{}^{\! |k|} e^{i\theta\hat n} & \text{ for } k \ge 0 
  \end{array}\right. ,
\end{equation}
where $\theta \in[-\pi,\pi)$. A negative $k$ thus denotes a downwards shift in boson number and a positive $k$ an upwards shift. This error basis $\hat E_k(\theta)$ is composed of rotations (parameterized by $\theta$ and generated by $\hat n$) and shifts (parameterized by $k$ generated by $\hat a^{k}/\hat a^{\dagger k}$).

The most physically relevant noise channel for bosonic hardware is the loss channel. The Kraus operators for the loss channel naturally described by powers of the annihilation operator $\hat a$ and thus these Kraus operators can be easily expressed in the basis \eqref{eq:E}. Unfortunately, that basis is not ideally suited to rotation codes, as they are designed to correct shifts in photon number without the $(\sqrt{n})^k$ factors that come from powers of the lowering operator. 

Inspired by the basis above, the authors of Ref.~\cite{CombesAlbertetal2021} define a new one that is more naturally suited to the errors corrected by rotation codes. As the rotation codes we consider are intimately related to the Susskind-Glogower phase operators, this new basis replaces the annihilation operator in \cref{eq:errormodel,eq:E} with powers of $\phaseOP_1^-$. Specifically, the new error basis is
\begin{equation}\label{eq:naturalerrormodel}
    \left\{ e^{i\theta\hat n}\phaseOP_k^-,\, \phaseOP_k^{+} e^{-i\theta \hat n} \right\},
\end{equation}
where $\phaseOP_k^- = (\phaseOP_1^-)^k$ as in \cref{eq:pow_susskind}. For convenience we denote the elements of this basis with a single symbol
\begin{equation}\label{eq:new_err_basis}
  \E{k}{\theta} \equiv \left\{ \begin{array}{ll}
      e^{i\theta\hat n} \phaseOP_k^- & \text{ for } k < 0 \\
      \phaseOP_k^+ e^{i\theta\hat n} & \text{ for } k \ge 0 
  \end{array}\right. .
\end{equation}
where $k\in \mathbb{Z}$ is the number of shift down (loss like) or shift up (gain like) events and $\theta \in [-\pi, \pi)$ is a rotation. These operators form an (overcomplete) basis for the bosonic Hilbert space, as proved in Ref.~\cite{CombesAlbertetal2021}, and thus one can decompose any quantum operation into this basis.

%============================================
\section{Error Propagation}\label{sec:errprop}
%============================================

In this section we see how errors at the input of a gate propagate (or commute) through the gate using the error basis introduced in \cref{eq:new_err_basis}. We are primarily interested in whether the operation of a gate produces additional errors when an error is present at its input. If additional errors are introduced by such error propagation we say that the error has been ``amplified''. Provided the errors are not amplified too much it should still be possible to build fault tolerant error correcting schemes. For multiqubit gates an error may spread to other qubits in addition to the error amplification on the original qubit. Several of the results in this section were first derived in Ref.~\cite{GrimCombBara20} in the error basis \eqref{eq:E}. We repeat those calculations in the new error basis \eqref{eq:new_err_basis} and derive new results as well. \cref{table:gate_comparison} summarizes the relationship between the prior work and our results. All error propagation formulas have been derived analytically and verified numerically.

\begin{table}[h!]
\centering  
\begin{tabular}{|c|c|c|}
\hline
Gate & GCB & This work \\
\hline
$\ZN$  & \checkmark   & r \\\hline
$\XN$  & \ding{55}    & \checkmark \\\hline
$\XN'$ &   \ding{55}  & \checkmark \\\hline
$\SN$  & \checkmark  & r \\\hline
$\TN$  & \checkmark  & r \\ \hline
$\TN'$ &  \ding{55}   & \checkmark  \\\hline
$\hat{R}_N(\phi_\ell)$ &  \ding{55}   & \checkmark  \\ \hline
$\hat{R}_N'(\phi_\ell)$ &  \ding{55}   & \checkmark  \\ \hline
$\hat{P}_N(\phi)$ &  \ding{55}   & \checkmark  \\\hline
CROT & \checkmark  & r \\\hline
CCROT &  \ding{55}   & \checkmark  \\ 
\hline
\end{tabular}
\caption{Comparison of error propagation in gate implementations as first derived by Grimsmo, Combes, and Baragiola (GCB)~\cite{GrimCombBara20} and as presented in this paper. A `$\checkmark$' indicates the formula originated in the respective paper, `\ding{55}' denotes it was not considered, and `r' signifies it was rederived in a new basis. Two gates in this table have not been introduced yet. The $\hat{R}_N$ gates are rotations about $\Z$ by a discrete angle $\phi_\ell = \pi/2^\ell$. The gate $\hat{P}_N$ is an arbitrary rotation about $\Z$.}
\label{table:gate_comparison}
\end{table}

%============================================
\subsection{General error propagation}
%============================================

Consider the circuit $\hat{G} \E{k}{\theta}$ which represents the occurrence of an arbitrary error $\E{k}{\theta}$ prior to the action of a unitary gate $\hat{G}$. Our goal is to find an equivalent circuit that appears as if an error (not necessarily the same one) occurred {\em after} the gate was applied: $\hat{G} \E{k}{\theta} = \hat{E}' \hat{G}$. We call this error propagation. The propagation of an arbitrary error can be determined through conjugation
\begin{equation} \label{eq:errprop}
    \hat{G} \E{k}{\theta} = \hat{G} \E{k}{\theta} \hat{G}^{\dagger} \hat{G} = \hat{E}' \hat{G}
\end{equation}
where $\hat{E}' \equiv \hat{G} \E{k}{\theta} \hat{G}^{\dagger}$. We will see that in many cases, we can decompose $\hat{E}' = \hat{F} \E{k'}{\theta'}$, i.e. there will be an error $\E{k'}{\theta'}$ that is similar in form to the original error, and there could be additional errors denoted here by $\hat{F}$. The corresponding quantum circuit representation of this process is
\begin{center}
\begin{quantikz}[row sep=0.295cm,column sep=0.275cm]
    & \gate{\E{k}{\theta}} & \gate[][0.7cm][0.7cm]{\hat G} & \qw
\end{quantikz}
\hspace{0.2em} \raisebox{-0.25em}{=} 
\begin{quantikz}[row sep=0.295cm,column sep=0.275cm]
    & \gate[][0.7cm][0.7cm]{\hat G} & \gate{\E{k'}{\theta'}} & \gate[][0.7cm][0.7cm]{\hat F} &\qw
\end{quantikz}.
\end{center}
The above process is the simplest description of the kinds of manipulations we will encounter below.

Recall, the error basis $\E{k}{\theta}$ is composed of rotations (parameterized by $\theta$ and generated by $\hat n$) and shifts (parameterized by $k$ generated by $\phaseOP_k^{\pm}$). In the context of our bosonic codes, most of the gates we consider are exponential functions of polynomials in $\hat n$, which are easily seen to commute with the rotation errors. To work out how the shift errors propagate through such gates, we see that for a well-behaved function $f$ of only the operator $\hat{n}$,
\begin{equation} \label{eq:general_function_gate_prop}
    e^{i f(\hat{n})} \E{k}{\theta} = e^{i[f(\hat{n}) - f(\hat{n}-k\hat{I})]} \E{k}{\theta} e^{if(\hat{n})}.
\end{equation}
Furthermore, we note
\begin{equation} \label{eq:linear_error_modification}
    e^{i\phi \hat{n}} \E{k}{\theta} = e^{i\phi k \Theta(k)}\E{k}{\theta + \phi}
\end{equation}
where $\Theta(x)$ is the Heaviside step function. We use these relations, derived in \cref{sec:derivations_appendix}, liberally in the following sections.

%============================================
\subsection{Single qubit gates} \label{sec:singlequbitgates}
%============================================

In this subsection we examine the propagation of an arbitrary single qubit error through the gates in \cref{eq:singlequbitgates}. Starting with $\ZN$, we find
\begin{equation} \label{eq:Z_spread}
    \ZN \E{k}{\theta} = e^{i \frac{k \pi}{N}} \E{k}{\theta} \ZN
\end{equation}
which holds for all $k \in \mathbb{Z}$. Error propagation through $\ZN$ only introduces a global phase -- the error is not amplified. As the $\ZN$ gate commutes with rotation errors since they are both functions of $\hat{n}$, the global phase in \cref{eq:Z_spread} is a function of $k$ alone.

When the physical gate is not unitary, as is the case for $\XN$, then the error propagation formula in \cref{eq:errprop} requires a little more thought. We need to be mindful of the fact that $\XN (\XN)^\dagger = \hat I$ but $(\XN)^\dagger \XN = \hat I - \hat{ \mathbb{P}}_N$ where $\hat{\mathbb{P}}_N = \sum_{n=0}^{N-1}\op{n}{n}$ is the projector onto the first $N$ Fock states. The propagation is still straightforward:
\begin{equation}\label{eq:X_spread}
   \XN \E{k}{\theta} = e^{i\theta N} \E{k}{\theta} \XN + e^{i \theta(N-k) \Theta(N-k)} \hat{\mathbb{P}}_k \E{k-N}{\theta}.
\end{equation}
For $k<0$ the global phase is only a function of $\theta$ since the shifting action of $\XN$ commutes with any shift errors.  Interestingly for $k\ge 0$, the first term on the RHS does not have support on the first $k$ Fock states while the LHS does. This means that errors through the $\XN$ gate are not recoverable unless the code begins without support on the first $k$ Fock states.  This feature will be important when we introduce our explicit error correction scheme in \cref{sec:explicit_QEC}.

Error propagation for the $\XN'$ gate is more complicated and can be written as follows
\begin{equation} \label{eq:XPrime_spread}
    \XN' \E{k}{\theta} = \left[ \sum_{\ell = 0}^{N-1} e^{i \theta p^+} \tilde{\mathbb{E}}(\theta)_{k - x_\ell}^\ell + e^{i \theta p^-} \tilde{\mathbb{E}}(\theta)_{k + x_\ell}^{\ell + N} \right] \XN'.
\end{equation}
Here we define
\begin{equation*}
    x_\ell \equiv \left\{ \begin{array}{ll} 2N & \text{if } \max(k\texttt{\%} 2N -N, 0) \le \ell < \min(k\texttt{\%} 2N, N) \\ 0 & \text{otherwise} \end{array} \right.
\end{equation*}
where \texttt{\%} represents the modulo operator, and
\begin{equation*}
    p^\pm \equiv \pm N- k\Theta(k) + (k \mp x_\ell)\Theta(k \mp x_\ell).
\end{equation*}
Additionally, $\tilde{\mathbb{E}}(\theta)_k^m$ is the error $\E{k}{\theta}$ projected onto the Fock states that are equivalent to $m \mod 2N$,
\begin{align}
\begin{aligned} \label{eq:modular_error}
    \tilde{\mathbb{E}}(\theta)_k^m &\equiv \sum_{n=0}^\infty \op{2nN+m}{2nN+m} \E{k}{\theta} \\
    &= \frac{1}{2N} \sum_{j=0}^{2N-1} e^{-i \frac{\pi j}{N} (m-k\Theta(k))} \E{k}{\theta + \frac{\pi j}{N}}.
\end{aligned}
\end{align}
The algebraic equivalence in the second line resembles a Fourier transform, and makes it clear that the bracketed term in \eqref{eq:XPrime_spread} can be decomposed as a linear combination of errors from our error basis. Alternate representations of $\X$ as a permutation of Fock states (e.g. we could permute the $\ket{kN}$ states as in \eqref{eq:X_prime_gate}, leaving all other Fock states undisturbed) will likely have similar error expressions.

For error propagation through the $\SN$ gate, we find
\begin{equation} \label{eq:S_prop}
    \SN \E{k}{\theta} = e^{i\phi_S} \E{k}{\theta+\frac{\pi k}{N^2}} \SN,
\end{equation}
where $\phi_S \equiv \frac{\pi k^2}{N^2}(\Theta(k)-\frac{1}{2})$. Again we see the introduction of a global phase, but now the rotation error parameter $\theta'=\theta+ \pi k /N^2$ is a linear function of the shift $k$. This becomes an issue if the original angular error $\theta$ was correctable but $\theta'$ is not, which happens when $k$ and $\theta$ are sufficiently large.

To conclude this section we will look at error propagation through two implementations of the $T$ gate, so that we cover a gate set sufficient for universal computation. Looking at the $\ZN$ and $\SN$ gate implementations, we expect that implementations of $T$, which is an even smaller $\Z$ rotation, will require a higher order polynomial in $\hat{n}$ in its exponential form, which will give rise to more significant error amplification. For the $\TN$ gate, error propagation yields
\begin{equation} \label{eq:T_prop}
    \TN \E{k}{\theta} = e^{i\phi_T} \hat{F}_N(k)\, \E{k}{\theta + \frac{\pi k^3}{N^4}} \TN,
\end{equation}
where $\phi_T \equiv \frac{\pi k^4}{N^4}(\Theta(k)-\frac{1}{4})$ and $\hat{F}_N(k) \equiv \exp \left[ i\frac{\pi}{4N^4} (4 k \hat{n}^3 - 6k^2 \hat{n}^2) \right]$ is a nonlinear rotation error. As with the $\SN$ gate, error propagation introduces a global phase and modifies the linear rotation error parameter $\theta$, but the additional error $\hat{F}_N(k)$ makes the propagated error much more complicated.

The $\TN'$ gate error propagation is a bit tamer because the order of $\hat{n}$ in its implementation is lower than that of $\TN$:
\begin{equation} \label{eq:T_prime_prop}
	\TN' \E{k}{\theta} = e^{i\phi_T'} \hat{F}_N'(k) \E{k}{\theta + \left( \frac{\pi k}{2N^2} - \frac{3\pi k^2}{2N^3} \right)} \TN'
\end{equation}
where $\phi_T' \equiv \frac{\pi}{4}\left[ \frac{2k^3}{N^3} - \frac{k^2}{N^2} -\frac{2k}{N} + \left( \frac{2k^2}{N^2} - \frac{6k^3}{N^3} \right) \Theta(k) \right]$ and the nonlinear error $\hat{F}_N'(k) \equiv e^{i \frac{3\pi k}{2N^3} \hat{n}^2}$.\footnote{ Note that the nonlinear term $\hat{F}_N'(k)$ is equivalent to $(\SN)^{3k/N}$, but it is not particularly illuminating to write it this way since in general it is not a logical operation on the codespace.} While seemingly more complicated, this implementation of the gate yields an error that is only quadratic in $\hat{n}$ compared with \cref{eq:T_prop} which has a cubic error in $\hat{n}$. This suggests that it may be a more useful implementation that allows more errors to be correctable.

%============================================
\subsection{Discrete and continuous rotation gates} \label{sec:rotation_gates}
%============================================

Now we will construct gates that perform arbitrary discrete and continuous $\Z$-rotations, which could be implemented using a SNAP gate~\cite{SNAPgateTHY2015,SNAPgateEXP2015}. The intuition for the constructions is essentially extending the pattern seen in the implementations of $\ZN$, $\SN$, and $\TN$ where smaller rotation angles required larger powers of $\hat n$. Unfortunately these smaller angle gates had errors that amplified in proportion to powers of $\hat n$ used to implement the gate. We show that this is a generic feature.

We first note that a rotation of $\phi_\ell = \pi/2^\ell$ can be achieved for $\ell \geq 0$ using 
\begin{equation} \label{eq:R_l_gate}
    \hat{R}_N(\phi_\ell) = \exp \left[ i \phi_\ell (\hat{n}/N)^{2^\ell} \right],
\end{equation}
which is consistent with the $\ZN, \SN$ and $\TN$ gates \cref{eq:Z_gatey,eq:S_gate,eq:T_gate} for $\ell=0,1,2$ respectively. The error propagation for this set of gates can easily be determined using \cref{eq:general_function_gate_prop,eq:linear_error_modification}:
\begin{align}
\begin{aligned}
    \hat{R}_N(&\phi_\ell) \E{k}{\theta} \\
    &= \exp\! \left[ \frac{i\pi}{2^\ell N^{2^\ell}} \left( \hat{n}^{2^\ell}-(\hat{n}-k)^{2^\ell} \right) \right] \E{k}{\theta} \hat{R}_N(\phi_\ell) \\
    &= \exp\! \left [i \mathcal{O} \left( \hat{n}^{2^\ell-1} \right) \right ] \mathbb E_{k}\!\left (\theta + \frac{\pi}{k} \left( \frac{k}{N} \right)^{\! 2^\ell} \right) \hat{R}_N(\phi_\ell).
\end{aligned}
\end{align}
The induced nonlinear error is of order $\hat{n}^{2^\ell -1}$ and the modification to the linear rotation error is of order $k^{2^\ell -1}$, which in general is quite bad. We can see that this error propagation formula has powers of $k$ that are consistent with those found in \cref{eq:Z_spread,eq:S_prop,eq:T_prop}.

Motivated by the $\TN'$ gate~\cite{AlvCalcFerrFerr20}, which provided an alternative implementation to $\TN$ with less significant error amplification, we propose an alternative set of discrete $Z$-rotation gates that lessens the propagated error. As seen in \cref{sec:rotation_implementation}, the optimal such implementation is of the form
\begin{equation}
    \hat{R}_N'(\phi_\ell) = \exp \left[ i \mathcal{O} \left( \hat{n}^{\ell + 1} \right) \right],
\end{equation}
which is consistent with the $\TN'$ gate \cref{eq:T_gate_prime}. Errors propagate through these gates like
\begin{equation}
    \hat{R}_N'(\phi_\ell) \E{k}{\theta} = e^{i \mathcal{O}(\hat{n}^\ell)} \E{k}{\theta + \mathcal{O}(k^\ell)} \hat{R}_N'(\phi_\ell)
\end{equation}
where now the induced nonlinear error is of order $\hat{n}^\ell$ and the modification to the linear rotation error is $\pi k^\ell/N^{\ell +1}$. This is a significant improvement over the initial implementation, although the errors are still quite bad in general. Again the power of $k$ in the expression above is consistent with \cref{eq:T_prime_prop}.

A method to construct a gate set that performs arbitrary continuous $Z$-rotations is to find a function $f(n)$ such that $f(kN) = 0$ for $k$ even and 1 for $k$ odd. Then the gate $e^{i \phi f(\hat{n})}$ will implement a $Z$-rotation by angle $\phi$ as desired. One simple function that satisfies this criteria is $f(n) = (1-\ZN)/2$, where we recall that $\ZN$ is itself a function of $\hat{n}$. Thus a general rotation gate is realized as
\begin{equation}
    \hat{P}_N(\phi) = \exp\left [i\frac{\phi}{2}(1-\ZN)\right ].
\end{equation}
We then find by the usual technique that
\begin{equation} \label{eq:phase_gate}
    \hat{P}_N(\phi) \E{k}{\theta} = \exp\! \left[ i\frac{\phi}{2} \left( e^{-i\frac{\pi k}{N}} - 1 \right) \ZN \right] \E{k}{\theta} \hat{P}_N(\phi)
\end{equation}
which induces a worse nonlinear error than the discrete rotation, as we would naively expect. 
Based on this gate we conjecture that a logical $\CPHASE(\phi)$ gate for arbitrary $\phi$ would involve a generator proportional to $i\phi (1-\ZN )\otimes (1-\ZN)$.\\

%============================================
\subsection{Multiqubit gates}
%============================================

Multiqubit gates may allow errors on one qubit to spread to other qubits. In the case of the $\CROT$ gate, an arbitrary error on the first qubit propagates through that mode unchanged while also introducing a pure rotation error on the second qubit. Specifically we find 
\begin{equation}\label{eq:crot_prop}
    \CROT_{NM}  \E{k}{\theta} \otimes \hat{I}  =  \E{k}{\theta} \otimes \E{0}{\frac{\pi k}{NM}}  \CROT_{NM} \, ,
\end{equation}
which is analogous to the result in Ref.~\cite{GrimCombBara20}. By symmetry, we can see than when the input error is on the second qubit instead, error propagation yields an identical expression for the output error, acting on the opposite qubits. Furthermore, these two amplified errors must commute, so if both input qubits are corrupted we may simply concatenate propagated errors of the form in \cref{eq:crot_prop} acting on the two qubit orderings:
\begin{widetext}
\begin{align}
\begin{aligned}
    \CROT_{NM} &\left( \E{k_1}{\theta_1} \otimes \E{k_2}{\theta_2} \right) = \left( \E{k_1}{\theta_1} \E{0}{\frac{\pi k_2}{NM}} \otimes \E{0}{\frac{\pi k_1}{NM}} \E{k_2}{\theta_2} \right) \CROT_{NM} \\
    &= \exp \left[ -i \frac{\pi k_1 k_2}{NM}(\Theta(-k_1) +\Theta(k_2)) \right] \left( \E{k_1}{\theta_1 + \frac{\pi k_2}{NM}} \otimes \E{k_2}{\theta_2 + \frac{\pi k_1}{NM}} \right) \CROT_{NM}.
\end{aligned}
\end{align}
\end{widetext}

A similar analysis follows for the CCROT gate, which acts as a logical $\CCZ$. In that case, the error propagation expression for an input error on a single qubit is
\begin{equation} \label{eq:ccrot_prop}
    \CCROT_{NMO} \E{k}{\theta} \otimes \hat{I}^{\otimes 2} = \E{k}{\theta} \otimes \hat{V}_k \,\CCROT_{NMO}
\end{equation}
where
\begin{equation}
    \hat{V}_k = \exp\left[ i \frac{\pi k}{NMO} \hat{n} \otimes \hat{n} \right].
\end{equation}
Again the input error propagates through unchanged on the affected qubit, while a new error $\hat V_k$\footnote{We can see that $\hat V_k = (\CROT_{MO})^{k/N}$, but as in the previous footnote this is not a particularly illuminating representation.} is introduced that acts on the other two qubits, shown in the following circuit diagram
\begin{center}
\begin{quantikz}[row sep=0.35cm,column sep=0.25cm]
    & \gate{\E{k}{\theta}} & \gate[wires=3, label style={rotate=-90}][0][0.7cm]{\rm CCROT} & \qw \\
    & \qw & \ghost[2]{} & \qw \\
    & \qw &  & \qw
\end{quantikz}
\hspace{0.2em} \raisebox{-0.3em}{=}
\begin{quantikz}[row sep=0.3cm,column sep=0.25cm]
    & \gate[wires=3,label style={rotate=-90}][0][0.7cm]{\rm CCROT} & \gate{\E{k}{\theta}} & \qw \\
    & & \gate[wires=2][1.2cm][0.7cm]{\hat V_k} & \qw \\
    & & & \qw
\end{quantikz}.
\end{center}
As in the $\CROT$ case, an input error on either of the other two qubits will behave symmetrically, and again the output error will commute with that of expression above. Thus if any subset of the input qubits are corrupted, the propagated error is given by concatenation of output errors of the form \cref{eq:ccrot_prop} acting on the relevant qubit permutations.

%============================================
\subsection{Comparison with qubit error propagation}
%============================================

To build intuition for the above expressions we derive analogous expressions for qubits. One motivation for this is that it is tempting to analyze errors on continuous variable codes in a logical qubit representation. This would provide a two dimensional description of continuous variable codes living in an infinite dimensional Hilbert space. It is an open question if such a representation can faithfully capture continuous variable errors after repeated rounds of error correction. 

To make a fair comparison between the two situations we must introduce a new error basis for qubits that is analogous to the bosonic error basis defined in \cref{eq:new_err_basis}. We define the qubit error basis as
\begin{equation}\label{eq:qubit_err_basis}
    \hat Q_{k}(\theta)\equiv \left\{ \begin{array}{ll}
        e^{i\theta\hat N} \sigma_- = \sigma_- & \text{ for } k =-1 \\
        e^{i\theta\hat N} & \text{ for } k = \, 0 \\
        \sigma_+ e^{i\theta \hat N} = \sigma_+ & \text{ for } k = +1
    \end{array}\right. ,
\end{equation}
where the qubit raising and lowering operators
\begin{equation}
    \sigma_\pm = (X \mp iY)/2
\end{equation}
are labeled oppositely from the usual qubit convention so that they raise and lower the numeric label of the state, analogous to the bosonic operators. Here we have $\sigma_- = \op{0}{1}$, $\sigma_+ = \op{1}{0}$, and $\hat N = \sigma_+ \sigma_- = \op{1}{1}$. These three operators have the same matrix elements as the bosonic $\hat{\Sigma}_1^-,\, \hat{\Sigma}_1^+,\, \hat n$ respectively, restricted to the subspace spanned by the vacuum and a single boson. It is simple to show that the operators $\hat Q_{k}(\theta)$ are a valid qubit basis by forming the Pauli matrices from linear combinations of the set of operators.

It is straightforward to commute the error basis through the single qubit gates analogous to those examined in \cref{sec:errprop}. Doing so gives
\begin{subequations} \label{eq:qubit_error_prop_1Q_gates}
\begin{align}
    \hat Z \hat Q_k (\theta) &= e^{ik \pi} \hat Q_{k}(\theta)\hat Z \,, \\
    \hat X \hat Q_k (\theta) &= e^{i\theta \delta_{k,0}} \hat Q_{-k} (-\theta) \hat X \,, \\
    \hat S \hat Q_k (\theta) &= e^{ik \pi/2}\hat Q_k(\theta) \hat S \,, \\
    \hat T \hat Q_k (\theta) &= e^{ik \pi/4}\hat Q_k(\theta) \hat T \,.
\end{align}
\end{subequations}
If we compare these expressions to the bosonic ones derived above by taking $N=1$ and restricting to the vacuum and single boson subspace, all error propagation formulae match. Note that the qubit gate $\hat{X}$ corresponds to the bosonic $\XN'$ rather than $\XN$ (the qubit analog of $\XN$ is simply $\tilde X=\sigma_-$, and error propagation through this gate matches the bosonic formula for $\XN$). 

Similarly, for the multiqubit gates we find
\begin{subequations} \label{eq:qubit_error_prop_multiQ_gates}
\begin{align}
    \CZ(\hat{Q}_k(\theta) \otimes \hat{I}) &= (\hat{Q}_k(\theta) \otimes \hat{Q}_0 (k \pi)) \CZ \\
    \CCZ(\hat{Q}_k(\theta) \otimes \hat{I}^{\otimes 2}) &=  (\hat{Q}_k(\theta) \otimes e^{-ik\pi \hat N \otimes \hat N}) \CCZ.
\end{align}
\end{subequations}
The form of the qubit error propagation looks identical to that of the bosonic errors for the $\hat{Z},\, \CZ$, and $\CCZ$ gates if we take $N=1$ (and $M=O=1$ as well). We note that while the expressions for the qubit error propagation are equivalent to the full bosonic expressions on the restricted subspace (the $0$ and $1$ Fock states), this doesn't imply that a general bosonic error projected into the logical subspace will be anything like the original error.

%============================================
\section{Explicit Error Correction} \label{sec:explicit_QEC}
%============================================

In this section we consider two explicit error correction schemes. In both cases we apply the $\Z$ and $\X$ stabilizers to diagnose an arbitrary error, followed by an error recovery operator. The first scheme applies the stabilizers directly to a corrupted state, while the second uses a teleportation scheme. The advantage of the latter scheme is that we do not lose any information when applying the $\X$ stabilizer, as explained below.

In the formulation of rotation codes we gave in \cref{sec:summary}, codewords may have support on the vacuum state. Thus if a single loss event occurs, the amplitude of vacuum is lost irrevocably. In  general this is not an issue for recovering rotation codes after such an error. Indeed, a single loss error can be recovered even in small binomial codes~\cite[see Sec. II]{Michael16}. However the recovery scheme we propose below is naive, limited to shifting amplitudes in Fock space and rotating the code in phase space. Therefore, to accommodate shift down errors of up to $N$, the code must start at the state $\ket{N}$ and not $\ket{0}$. Furthermore, the $\X$ stabilizer induces an additional shift downwards by $2N$. Thus in general to avoid any loss of information from the error recovery procedure we must start the code at $3N$, i.e. codewords must not have support on states $\ket{m}$ for $m<3N$. This is the case in \cref{sec:directstab}. 

When the $\X$ stabilizer is not applied directly to the state we wish to recover, we may instead start the code at $N$ as we do in \cref{sec:knill}. However, it should be noted that if the code is an approximate phase code, as it must be if \cref{eq:X_gate} is to approximate an X gate, this caveat becomes irrelevant and we need not amend our codewords in either case.

As a general description of our potentially modified codewords, we shift the starting point of the Fock grid in \cref{eq:logicalz_codewords} and \cref{eq:logicalx} to the state $\ket{k_0 N}$:
\begin{subequations}\label{eq:codeword_offset}
\begin{align}
    \ket{\zeroL'} &= \sum_{k=k_0}^{\infty} f_{2kN}\ket{2kN} \\
    \ket{\oneL'} &= \sum_{k=k_0}^{\infty} f_{(2k+1)N}\ket{(2k+1)N} \\
    \ket{\plusL'} &=  \frac{1}{\sqrt{2}}\sum_{k=k_0}^\infty  f_{kN} \ket{kN} \\
    \ket{\minusL'} &=  \frac{1}{\sqrt{2}} \sum_{k=k_0}^\infty  (-1)^k f_{kN} \ket{kN}.
\end{align}
\end{subequations}
When $k_0=0$ we have the original codeword definitions (see \cref{eq:logicalz_codewords,eq:logicalx}), while $k_0=3$ and $k_0=1$ refer to the codewords necessary for general error recovery under the two correction schemes. Note that $\ket{\zeroL'} = \phaseOP_{k_0 N}^+ \ket{\zeroL}$ and similarly for the other codewords.

%============================================
\subsection{Direct stabilizer measurement}\label{sec:directstab}
%============================================

In this scheme we assume it is possible to directly, and non-destructively, measure stabilizers. Here we take $k_0 = 3$. The error correction scheme depicted below has an error $\E{k}{\theta}$ followed by the two stabilizer measurements from \cref{eq:stabilizers}. To show that this works as an error correction scheme we will apply all of the circuit elements to an encoded quantum state and see that the circuit output is the same as the input $\ket{\psi}$ for a suitable error recovery operator $\hat{\mathcal{R}}$.
\begin{equation} \label{eq:recovery_circuit}
\begin{quantikz}[row sep=0.4cm,column sep=0.5cm]
    &\lstick{$\ket{\psi}$} &\gate{\E{k}{\theta}} &\meterD{\hat S_{\Z}} &\meterD{\hat S_{\X}} &\gate[][0.7cm][0.7cm]{\hat{\mathcal{R}}} &\qw \\
    & &\cw &\cwbend{-1} &\cw &\cwbend{-1} &\cw \\
    & &\cw &\cw &\cwbend{-2} &\cwbend{-1} &\cw  
\end{quantikz} \,.
\end{equation}
Of course the error element $\E{k}{\theta}$ is not a quantum channel. However we can decompose the Kraus operators of the quantum channel into our error basis, this motivates considering an error from the basis directly.

For the manipulations below it is convenient express the encoded state in the dual basis. An arbitrary initial state is then given by
\begin{equation}
    \ket{\psi} = \alpha \ket{\plusL'} + \beta \ket{\minusL'} \,. 
\end{equation}
An arbitrary element of our error basis \cref{eq:new_err_basis} acting on the encoded state is then
\begin{align} \label{eq:psi_prime}
\begin{aligned}
    \ket{\psi_E} &= \,\hat{\mathbb{E}}_m(\theta)\ket{\psi} \\
    &= \frac{e^{i \theta m \Theta(-m)}}{\sqrt{2}} \sum_{k=k_0}^{\infty} (\alpha + (-1)^k \beta) e^{i \theta kN} f_{kN} \ket{kN+m}, 
\end{aligned}
\end{align}
and where $\Theta(x)$ is the Heaviside step function.

Next we examine the action of the stabilizers on the state $\ket{\psi_E}$. The $\hat{S}_{\Z}$ stabilizer will allow us to determine the error parameter $m$ while $\hat{S}_{\X}$ will determine the error parameter $\theta$, and since the stabilizers commute the order of application does not matter. Looking at the action of the stabilizers on an error state one at a time, we see
\begin{equation} \label{eq:S_Z_signal}
    \hat{S}_{\Z} \ket{\psi_E} = \exp\! \left ( \frac{2 \pi i }{N} m \right ) \ket{\psi_E} \equiv \lambda_Z(m) \ket{\psi_E}.
\end{equation}
Evidently the number stabilizer $\hat{S}_{\Z}$ produces an error signal that is a function $m$, without introducing additional errors to the state $\ket{\psi_E}$. Similarly,
\begin{widetext}
\begin{equation}
    \hat{S}_{\X} \ket{\psi_E} = e^{i2N\theta} \frac{e^{i\theta m \Theta(-m)}}{\sqrt{2}} \sum_{k=k_0-2}^{\infty} (\alpha + (-1)^k \beta) e^{i\theta kN} f_{(k+2)N} \ket{kN+m} \equiv e^{i2N\theta}\ket{\psi_E'}
\end{equation}
\end{widetext}
which we can write more compactly as
\begin{equation} \label{eq:S_X_signal}
    \hat{S}_{\X} \ket{\psi_E} = e^{i2N\theta}\ket{\psi_E'} \equiv \lambda_X(\theta) \ket{\psi_E'},
\end{equation}
where $\ket{\psi_E'}$ only differs from $\ket{\psi_E}$ in the summation index and the $f$ amplitudes. The error signal $\lambda_X$ of this stabilizer can be extracted from a nondestructive measurement of modular phase, see e.g. \cite{GrimCombBara20}, from which we can extract the parameter $\theta$.

Applying both stabilizers, we see
\begin{equation} \label{eq:syndromes}
    \hat{S}_{\Z} \hat{S}_{\X} \ket{\psi_E} = \hat{S}_{\X} \hat{S}_{\Z} \ket{\psi_E} = \exp\! \left [ \frac{2\pi i}{N}m + i2N\theta \right ] \ket{\psi_E'}
\end{equation}
for any error state. Having determined $m$ and $\theta$ from the stabilizer measurements, our task is to restore $\ket{\psi_E'}$ to the original state $\ket{\psi}$ with a recovery operator $\hat{\mathcal{R}}_{m, \theta}$.

\begin{figure}
    \includegraphics[width=\columnwidth]{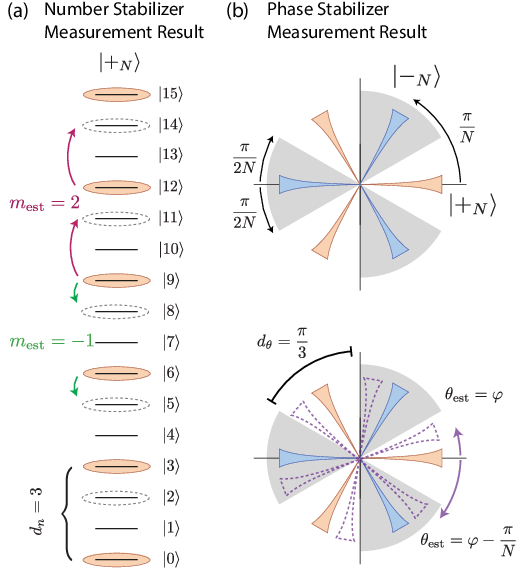}
    \caption{Error detection and recovery scheme. In (a), a truncated Fock grid is depicted for an $N=3$ code, where the codespace support states are on the ``Fock grid'' $\ket{Nk}$. A shift error $\E{m}{0}$ has occurred on $\ket{+_N}$ (where we take $k_0=0$ for diagrammatic convenience), and the resulting state has support depicted by the dotted ovals. The error could have either been a shift down by 1 or a shift up by 2; both will yield the same stabilizer syndrome. Thus the choice of the estimator $m_{\rm est}$ and hence the recovery operation should be chosen based on the corresponding error probabilities for a given noise model. In (b)  we depict the $N=3$ code in phase space, with $\ket{+_N}$ in orange and $\ket{-_N}$ in blue. The top illustration shows a logical $\ZN$ gate is a rotation by $\pi/N$. The bottom illustration shows a phase error $\E{0}{\theta}$ has occurred with resulting code space rotation shown by the purple dots. The error could have been a rotation by angle $\varphi$ or $\varphi - \pi/N$; again, both will yield the same syndrome and the recovery operation. Thus the estimator $\theta_{\rm est}$ should be chosen according to the noise model. Note that unlike the discrete shift error, phase errors are continuous and can only be corrected up to the resolution of the phase stabilizer measurement.}\label{fig:error_recovery}
\end{figure}

We note several important facts. First, the syndrome $\lambda_Z(m) = e^{\frac{2\pi i}{N}m}$ from the $\hat{S}_{\Z}$ stabilizer suggests a naive estimator $\bar{m} = \frac{N}{2\pi} \text{Arg} \left( \lambda_Z(m) \right)$ which is only accurate up to integer multiples of $N$, see \cref{fig:error_recovery}. A better estimator may be constructed as $m_{\text{est}} = \bar{m} + lN$ where the integer $l$ is chosen based on the nature of the error channel. If, for example, a channel only admits gain errors and larger gains occur with decreasing probability, we can choose $l$ that minimizes $m_{\text{est}} \ge 0$. Similarly, for a channel that only admits losses with a decreasing probability of more losses, we can choose $l$ that maximizes $m_{\text{est}} \le 0$. For a channel that admits both gain and loss errors with decreasing probability in the total number of gains or losses, we choose $l$ that minimizes $|m_{\text{est}}|$. For an approximate phase code, any choice of $l$ that gives $m_{\text{est}} \ge -k_0 N$ works equally well. 

The $\theta$ estimator has a similar ambiguity. The naive estimator $\bar{\theta} = \frac{1}{2N} \text{Arg} \left( \lambda_X(\theta) \right)$ is only accurate up to integer multiples of $\pi/N$, see \cref{fig:error_recovery}. On the other hand, the phase parameter $\theta$ itself has modularity $2\pi/N$ so we must pick a $\theta_{\text{est}} \in \{ \bar{\theta}, \bar{\theta}-\frac{\pi}{N} \}$. If larger phase errors occur with decreasing probability in our error channel, we can choose the option that minimizes the $|\theta_{\text{est}}|$ as the most likely.

Finally, we see that the stabilizer $\hat{S}_X$ introduces an additional loss of $2N$ which we must undo as well in the recovery operation (and for approximate phase codes the condition on $l$ is in fact $m_{\text{est}} - 2N \ge -k_0 N$). Putting all the pieces together, we define the recovery operation as
\begin{equation}
    \hat{\mathcal{R}}_{m,\theta} = \hat{\mathbb{E}}_{2N-m_{\text{est}}}(-\theta_{\text{est}}).
\end{equation}
One might be tempted to think the shift of photon number in this recovery is physically unreasonable, however \citet{Gertler21} have experimentally implemented such recovery operations.
In summary, we have defined a recovery operation $\hat{\mathcal{R}}$ such that
\begin{equation}
    \hat{\mathcal{R}}_{m, \theta} \hat{S}_{\Z} \hat{S}_{\X} \hat{\mathbb{E}}_{m}(\theta) \ket{\psi} = \ket{\psi}
\end{equation}
up to a known overall phase, so long as $m$ and $\theta$ satisfy the error correction criteria detailed in \cref{sec:num_phase_trade}.

%============================================
\subsection{Teleportation based error correction}\label{sec:knill}
%============================================
A teleportation based error correction scheme for rotation codes, along the lines suggested by Knill~\cite{Knill2005}, was proposed by \citet{GrimCombBara20}. It is easy enough to extend our explicit scheme to deal with teleportation based correction of Grimsmo {\em et al.} using the circuit
\begin{equation} \label{eq:teleport}
\begin{quantikz}[row sep=0.35cm,column sep=0.3cm]
    &\lstick{$\ket{\psi}$} &\gate{\E{k}{\theta}} &\gate[wires=2,label style={rotate=-90}][0][0.7cm]{\rm CROT} &\qw &\meterD{\hat S_{\X}} &\cwbend{1} \\
    &\lstick{$\ket{+_N}$} &\qw & &\gate[wires=2,label style={rotate=-90}][0][0.7cm]{\rm CROT} &\meterD{\hat S_{\X}} &\cwbend{1} \\
    &\lstick{$\ket{+_N}$} &\qw &\qw & &\qw &\gate[][0.7cm][0.7cm]{\hat{\mathcal{R}}}
\end{quantikz}\,.
\end{equation} \vspace{3pt}

In this scheme two encoded plus states are consumed during a teleportation through two CROT gates. The analysis assumes no errors happen in state preparation or during the teleportation scheme.

The two $\hat S_{\X}$ measurements in \cref{eq:teleport} end up giving the $\lambda_X$ and $\lambda_Z$ syndromes, as the teleportation through a CROT effectively performs a logical Hadamard on the output state, so the second $\hat S_{\X}$ measurement is actually a modular number measurement. 

As we are teleporting onto clean auxiliary system there is no additional error induced by measuring the stabilizers. This means that if the code starts at $k_0 = 1$ it should work with our naive correction. This suggests that with an appropriate auxiliary coupled measurement of the stabilizers the code could also start at $k_0=1$.

%============================================
\section{Code distance and Number-phase error trade off} \label{sec:num_phase_trade}
%============================================

In this section we establish limits on correctable errors for single mode bosonic codes with rotation symmetry. In general, errors on bosonic codes are not exactly correctable by the Knill-Laflame conditions stated below; rather, they are only approximately correctable~\cite{ApproxLeung97}. Approximate error correction for bosonic rotation codes has been previously examined by several authors~\cite{Michael16, albert2018, Yingkai2021}, we complement and extend those results in a new error basis.

Exact error correction criteria require that the Knill-Laflame conditions
\begin{equation}
    \hat{\mathbb{P}}_L \hat{E}_a^\dagger \hat{E}_b \hat{\mathbb{P}}_L = \alpha_{ab} \hat{\mathbb{P}}_L
\end{equation}
are satisfied~\cite{KnillLaflamme97}. Here $\hat{\mathbb{P}}_L = \op{0_L}{0_L} + \op{1_L}{1_L} = \op{+_L}{+_L} + \op{-_L}{-_L}$ is the projector onto the logical codespace, $\hat{E}_a$ and $\hat{E}_b$ are generic errors from the set of correctable errors, and $\alpha_{ab}$ is a Hermitian matrix in the error space. When these conditions are satisfied, the two errors are mutually distinguishable and correctable by some recovery procedure (e.g. the one outlined in section \ref{sec:explicit_QEC}). 

The codespace projector for rotation codes is
\begin{equation}
    \hat{\mathbb{P}}_L = \frac{1}{2}\sum_{r,s = 0}^\infty f_{rN} f_{sN}^* \left[ 1 + (-1)^{r+s} \right] \op{rN}{sN},
\end{equation}
and for an ideal phase code the $f$ amplitudes are constant and can be ignored. Using these conditions, we can evaluate which errors from the error basis \eqref{eq:new_err_basis} are mutually correctable. A bit of algebra shows that, with $\varphi \equiv \phi-\theta$,
\begin{widetext}
\begin{align}
\begin{aligned} \label{eq:KL_full}
    \hat{\mathbb{P}}_L \left(\E{j}{\theta} \right)^\dagger& \E{k}{\phi} \hat{\mathbb{P}}_L \\
    = \frac{1}{4} &\sum_{r,s,m'} f_{rN} f_{sN}^* f_{m'-j}^* f_{m'-k} e^{i\varphi m'} \left[ 1 + (-1)^{r+\frac{m'-j}{N}} + (-1)^{s+\frac{m'-k}{N}} + (-1)^{r+s+\frac{m'-j}{N}+\frac{m'-k}{N}} \right] \op{rN}{sN}
\end{aligned}
\end{align}
\end{widetext}
where $\{m'\}$ is the subset of integers such that both $\frac{m'-j}{N}$ and $\frac{m'-k}{N}$ are also integers. We see immediately that if $j \ne k \mod N$, $\{m'\} = \emptyset$; the sum evaluates to 0 and the Knill-Laflamme conditions are satisfied trivially.

In the case $j = k+N$, the above expression \eqref{eq:KL_full} will reduce to
\begin{equation} \label{eq:KL_Nshift}
    e^{i\varphi j} \sum_{r,s',m'} f_{rN}f^*_{s'N}f^*_{m'N} f_{(m'+1)N} e^{i\varphi Nm'} \op{rN}{s'N}
\end{equation}
where the sum is over the subset of $\mathbb{Z}^3$ such that $s' \ne r \mod 2$ and $m' = r \mod 2$. Due to the restricted sum, this will never be proportional to $\hat{\mathbb{P}}_L$ unless the sum over $m'$ evaluates to zero. In a contrived code it may be possible for this to be the case\footnote{For example, consider the code with unnormalized amplitudes $f_{mN} = 1,i,i,1,-1,i,i,-1,1,\dots$. When $\varphi=0$, the sum over $m'$ in \eqref{eq:KL_Nshift} vanishes both when $m'$ is even and odd. For this code, the errors $\E{k}{\theta}$ and $\E{k+N}{\theta}$ are mutually detectable so the gate $\XN$ cannot be a logical operation on the codespace and is not a suitable representation of $\X$. In fact, the condition that the sum over $m'$ vanishes for $\varphi=0$ is equivalent to $\bra{0_N} \XN \ket{1_N} = 0$, whereas for a logical $X$ gate we would have $\bra{0_N} \X \ket{1_N} = 1$. Ref.~\cite{PhaseEngcodes} has made similar observations.}, but for the examples found in the literature it will not. Specifically, for an ideal phase code where we can neglect the constant $f$ amplitudes, more algebra simplifies the expression to
\begin{equation}
    e^{i\varphi (j + N/2)} \left[ \cos(\varphi N/2) \hat{X}_L - \sin(\varphi N/2) \hat{Y}_L \right]
\end{equation}
where $\hat{X}_L = \op{0_N}{1_N} + \op{1_N}{0_N}$ and $\hat{Y}_L = -i\op{0_N}{1_N} + i\op{1_N}{0_N}$ are logical $X$ and $Y$ operators. We see that this will never vanish (or be proportional to $\hat{\mathbb{P}}_L$) for any value $\varphi$. We therefore conclude the intuitive result that $\E{j}{\theta}$ and $\E{j+N}{\phi}$ are not mutually correctable. In other words, a shift error of $N$ photons, which is a logical operation ($\XN$), is not detectable as an error. Combined with the previous result we may choose any set of N integers none of which are equivalent $\mod N$, as mutually correctable shift errors. Depending on the noise model of an error channel, the most convenient such sets would be $\{ 0,\ldots, N \}$ if only photon gain errors are likely, $\{-N,\ldots, 0 \}$ if only photon loss errors are likely, or $\{ \lfloor -N/ 2 \rfloor, \ldots,  \lfloor N/2 \rfloor  \}$ if both gains and losses are possible and errors with a smaller change in photon number are more likely (see the discussion in section \ref{sec:explicit_QEC}).

Turning our attention to the case $j=k$, equation \eqref{eq:KL_full} reduces to
\begin{equation}
    \frac{1}{2} e^{i\varphi j} \sum_{m=0}^\infty |f_{mN}|^2 \left[ \left(e^{i\varphi N}\right)^m \hat{\mathbb{P}}_L + \left(-e^{i\varphi N}\right)^m \hat{Z}_L \right]
\end{equation}
where $\hat{Z}_L = \op{0_N}{0_N} - \op{1_N}{1_N}$ is a logical $Z$ operator. Note that $\sum_{m=0}^\infty |f_{mN}|^2 (\cdot)$ is the expectation of $(\cdot)$ over the support of the codespace, so we can write the above expression as
\begin{equation}
    \frac{1}{2} e^{i\varphi j} \left[ \left\langle e^{i\varphi N \hat{n}} \right\rangle \hat{\mathbb{P}}_L + \left\langle e^{i(\varphi N + \pi) \hat{n}} \right\rangle \hat{Z}_L \right].
\end{equation}
This will only be proportional to the codespace projector if $\left\langle e^{i(\varphi N + \pi) \hat{n}} \right\rangle = 0$. For an ideal phase code, the expectation of this phase operator vanishes unless $\varphi = (2k+1)\pi/N$ for integer $k$; this tells us that we may choose any set of phase errors that are not equivalent $\mod \pi/N$. This again matches intuition, since a phase error of $\pi/N$ is a logical operation ($\ZN$) and should not be detectable as an error. As above, the convenient choice for a set of mutually correctable phase errors is generally $(-\frac{\pi}{2N}, \frac{\pi}{2N})$. It should be noted however that this discussion only applies to an ideal phase code. For any other code, it is not immediately obvious whether $\left\langle e^{i(\varphi N + \pi) \hat{n}} \right\rangle = 0$ for any value $\varphi$, and for most values it will not vanish. Thus for these codes, phase errors are not perfectly detectable and any error correction scheme must be approximate. Finally, we recall that the Knill-Laflamme conditions require that the proportionality factor $\alpha_{(j,\theta),(k,\phi)}$ be Hermitian in the error space. Since $\varphi = \phi - \theta$ appears only in complex exponentials and we are considering the case where $j=k$, we see that $\alpha_{(j, \theta),(j,\phi)} = \alpha^*_{(j, \phi),(j,\theta)}$ as desired, and the errors are indeed correctable. 

In summary, the set of errors $\big\{ \E{n}{\theta}\! \big\}$ are mutually detectable for an order-$N$ ideal phase code when e.g.
\begin{subequations} \label{eq:code_distance}
\begin{align}
 n &\in \{ 0, \ldots, N-1 \} \quad {\rm (number\ errors)} \\
\theta  &\in \left (-\frac{\pi}{2N}, \frac{\pi}{2N} \right ) \quad {\rm (phase\ errors)}.
\end{align}
\end{subequations}
As $N$ increases we are able to correct more number errors but fewer phase errors and we have a trade off that suggests $N$ should be selected specifically to construct the most robust error correcting code for a given noise model. This validates the conjectured~\cite{GrimCombBara20} number (shift) and rotation error distance of an ideal phase code to be 
\begin{subequations} \label{eq:num_phase_tradeoff}
\begin{align}
d_n &= N \quad &{\rm (number\ distance)} \\
d_\theta & = \pi/N \quad &{\rm (phase\ distance)}
\end{align}
\end{subequations}
such that the number-phase error trade off is 
\begin{equation}
    d_n d_\theta = \pi\,,
\end{equation}
 which was also shown in Ref.~\cite{Yingkai2021}. We can see that the recovery scheme detailed in \cref{sec:explicit_QEC} achieves this bound by noting that in \cref{eq:syndromes}, syndrome measurements are distinct for distinct errors in the ranges \cref{eq:code_distance}.

%==========================================
\section{Conclusions}\label{sec:conclusion}
%==========================================
The tool that enabled most of the results in this article was an error basis suited to codes with rotation symmetry. Using this basis we derived the error propagation formula for several gates, presented an explicit error correction scheme, and shown how to compute the code distance for rotation codes.

Regarding the code distance for rotation codes, it is useful reflect upon the corresponding scenario within qubit codes. Qubit stabilizer codes natively correct Pauli errors. Thus the distance of qubit codes is naturally described in the error basis $\{I, X, Y, Z\}$ and the distance is the number of those errors a code can correct. Despite the importance of amplitude damping errors in physical qubits, discussing the code distance of a stabilizer code in terms of its ability to correct amplitude damping errors is not standard practice.

Similarly, in \cref{sec:num_phase_trade} we have used the natural error basis of a rotation code to determine the code distance for rotation codes with respect to photon number shift errors and phase errors. Although the loss and dephasing channels are of importance for bosonic systems the loss and dephasing rates $(\kappa_l, \kappa_\phi)$ are not natural for defining code distance in rotation codes.

We see several areas that would be fruitful to explore in future work. One area for future work would be to extend error propagation to unitary gates like the ``eSWAP'' gate ~\cite{GaoLesCho2019} and dissipative gates such as those in Refs.~\cite{Mirrahimi_2014,Chamberland_2022,Gautier_2023}.
Another area would be to determine the distance for arbitrary finite energy rotation codes, either numerically or analytically. 

\,\newline
\noindent {\em Acknowledgments:} The authors acknowledge several helpful discussions with Victor Albert, Ben Baragiola, Arne Grimsmo, Akira Kyle, and Noah Lordi. BM, MB, and JC were supported by the National Science Foundation through a CAREER award EPMD-2240129 and Quantum Leap Challenge Institutes (QLCI) award OMA-2016244.

%============================================
%\input{main.bbl}
%\bibliographystyle{apsrev4-2} 
\bibliography{refs.bib}

%Control: production of eprint (1) enabled
\begin{thebibliography}{31}%
\makeatletter
\providecommand \@ifxundefined [1]{%
 \@ifx{#1\undefined}
}%
\providecommand \@ifnum [1]{%
 \ifnum #1\expandafter \@firstoftwo
 \else \expandafter \@secondoftwo
 \fi
}%
\providecommand \@ifx [1]{%
 \ifx #1\expandafter \@firstoftwo
 \else \expandafter \@secondoftwo
 \fi
}%
\providecommand \natexlab [1]{#1}%
\providecommand \enquote  [1]{``#1''}%
\providecommand \bibnamefont  [1]{#1}%
\providecommand \bibfnamefont [1]{#1}%
\providecommand \citenamefont [1]{#1}%
\providecommand \href@noop [0]{\@secondoftwo}%
\providecommand \href [0]{\begingroup \@sanitize@url \@href}%
\providecommand \@href[1]{\@@startlink{#1}\@@href}%
\providecommand \@@href[1]{\endgroup#1\@@endlink}%
\providecommand \@sanitize@url [0]{\catcode `\\12\catcode `\$12\catcode
  `\&12\catcode `\#12\catcode `\^12\catcode `\_12\catcode `\%12\relax}%
\providecommand \@@startlink[1]{}%
\providecommand \@@endlink[0]{}%
\providecommand \url  [0]{\begingroup\@sanitize@url \@url }%
\providecommand \@url [1]{\endgroup\@href {#1}{\urlprefix }}%
\providecommand \urlprefix  [0]{URL }%
\providecommand \Eprint [0]{\href }%
\providecommand \doibase [0]{https://doi.org/}%
\providecommand \selectlanguage [0]{\@gobble}%
\providecommand \bibinfo  [0]{\@secondoftwo}%
\providecommand \bibfield  [0]{\@secondoftwo}%
\providecommand \translation [1]{[#1]}%
\providecommand \BibitemOpen [0]{}%
\providecommand \bibitemStop [0]{}%
\providecommand \bibitemNoStop [0]{.\EOS\space}%
\providecommand \EOS [0]{\spacefactor3000\relax}%
\providecommand \BibitemShut  [1]{\csname bibitem#1\endcsname}%
\let\auto@bib@innerbib\@empty
%</preamble>
\bibitem [{\citenamefont {Grimsmo}\ \emph {et~al.}(2020)\citenamefont
  {Grimsmo}, \citenamefont {Combes},\ and\ \citenamefont
  {Baragiola}}]{GrimCombBara20}%
  \BibitemOpen
  \bibfield  {author} {\bibinfo {author} {\bibfnamefont {A.~L.}\ \bibnamefont
  {Grimsmo}}, \bibinfo {author} {\bibfnamefont {J.}~\bibnamefont {Combes}},\
  and\ \bibinfo {author} {\bibfnamefont {B.~Q.}\ \bibnamefont {Baragiola}},\
  }\bibfield  {title} {\bibinfo {title} {Quantum computing with
  rotation-symmetric bosonic codes},\ }\href
  {https://dx.doi.org/10.1103/PhysRevX.10.011058} {\bibfield  {journal}
  {\bibinfo  {journal} {Phys. Rev. X}\ }\textbf {\bibinfo {volume} {10}},\
  \bibinfo {pages} {011058} (\bibinfo {year} {2020})}\BibitemShut {NoStop}%
\bibitem [{\citenamefont {Xu}\ \emph {et~al.}(2023)\citenamefont {Xu},
  \citenamefont {Wang},\ and\ \citenamefont {Albert}}]{xu_clifford_2023}%
  \BibitemOpen
  \bibfield  {author} {\bibinfo {author} {\bibfnamefont {Y.}~\bibnamefont
  {Xu}}, \bibinfo {author} {\bibfnamefont {Y.}~\bibnamefont {Wang}},\ and\
  \bibinfo {author} {\bibfnamefont {V.~V.}\ \bibnamefont {Albert}},\ }\bibfield
   {title} {\bibinfo {title} {Clifford operations and homological codes for
  rotors and oscillators},\ }\href {https://doi.org/10.48550/arXiv.2311.07679}
  {\bibfield  {journal} {\bibinfo  {journal} {arXiv}\ ,\ \bibinfo {pages}
  {preprint arXiv:2311.07679}} (\bibinfo {year} {2023})}\BibitemShut {NoStop}%
\bibitem [{\citenamefont {Ofek}\ \emph {et~al.}(2016)\citenamefont {Ofek},
  \citenamefont {Petrenko}, \citenamefont {Heeres}, \citenamefont {Reinhold},
  \citenamefont {Leghtas}, \citenamefont {Vlastakis}, \citenamefont {Liu},
  \citenamefont {Frunzio}, \citenamefont {Girvin}, \citenamefont {Jiang},
  \citenamefont {Mirrahimi}, \citenamefont {Devoret},\ and\ \citenamefont
  {Schoelkopf}}]{Ofek}%
  \BibitemOpen
  \bibfield  {author} {\bibinfo {author} {\bibfnamefont {N.}~\bibnamefont
  {Ofek}}, \bibinfo {author} {\bibfnamefont {A.}~\bibnamefont {Petrenko}},
  \bibinfo {author} {\bibfnamefont {R.}~\bibnamefont {Heeres}}, \bibinfo
  {author} {\bibfnamefont {P.}~\bibnamefont {Reinhold}}, \bibinfo {author}
  {\bibfnamefont {Z.}~\bibnamefont {Leghtas}}, \bibinfo {author} {\bibfnamefont
  {B.}~\bibnamefont {Vlastakis}}, \bibinfo {author} {\bibfnamefont
  {Y.}~\bibnamefont {Liu}}, \bibinfo {author} {\bibfnamefont {L.}~\bibnamefont
  {Frunzio}}, \bibinfo {author} {\bibfnamefont {S.~M.}\ \bibnamefont {Girvin}},
  \bibinfo {author} {\bibfnamefont {L.}~\bibnamefont {Jiang}}, \bibinfo
  {author} {\bibfnamefont {M.}~\bibnamefont {Mirrahimi}}, \bibinfo {author}
  {\bibfnamefont {M.~H.}\ \bibnamefont {Devoret}},\ and\ \bibinfo {author}
  {\bibfnamefont {R.~J.}\ \bibnamefont {Schoelkopf}},\ }\bibfield  {title}
  {\bibinfo {title} {Extending the lifetime of a quantum bit with error
  correction in superconducting circuits},\ }\href
  {https://doi.org/10.1038/nature18949} {\bibfield  {journal} {\bibinfo
  {journal} {Nature}\ }\textbf {\bibinfo {volume} {536}},\ \bibinfo {pages}
  {441} (\bibinfo {year} {2016})}\BibitemShut {NoStop}%
\bibitem [{\citenamefont {Hu}\ \emph {et~al.}(2019)\citenamefont {Hu},
  \citenamefont {Ma}, \citenamefont {Cai}, \citenamefont {Mu}, \citenamefont
  {Xu}, \citenamefont {Wang}, \citenamefont {Wu}, \citenamefont {Wang},
  \citenamefont {Song}, \citenamefont {Zou}, \citenamefont {Girvin},
  \citenamefont {Duan},\ and\ \citenamefont {Sun}}]{LyanSun}%
  \BibitemOpen
  \bibfield  {author} {\bibinfo {author} {\bibfnamefont {L.}~\bibnamefont
  {Hu}}, \bibinfo {author} {\bibfnamefont {Y.}~\bibnamefont {Ma}}, \bibinfo
  {author} {\bibfnamefont {W.}~\bibnamefont {Cai}}, \bibinfo {author}
  {\bibfnamefont {X.}~\bibnamefont {Mu}}, \bibinfo {author} {\bibfnamefont
  {Y.}~\bibnamefont {Xu}}, \bibinfo {author} {\bibfnamefont {W.}~\bibnamefont
  {Wang}}, \bibinfo {author} {\bibfnamefont {Y.}~\bibnamefont {Wu}}, \bibinfo
  {author} {\bibfnamefont {H.}~\bibnamefont {Wang}}, \bibinfo {author}
  {\bibfnamefont {Y.~P.}\ \bibnamefont {Song}}, \bibinfo {author}
  {\bibfnamefont {C.~L.}\ \bibnamefont {Zou}}, \bibinfo {author} {\bibfnamefont
  {S.~M.}\ \bibnamefont {Girvin}}, \bibinfo {author} {\bibfnamefont {L.-M.}\
  \bibnamefont {Duan}},\ and\ \bibinfo {author} {\bibfnamefont
  {L.}~\bibnamefont {Sun}},\ }\bibfield  {title} {\bibinfo {title} {Quantum
  error correction and universal gate set operation on a binomial bosonic
  logical qubit},\ }\href {https://doi.org/10.1038/s41567-018-0414-3}
  {\bibfield  {journal} {\bibinfo  {journal} {Nature Physics}\ }\textbf
  {\bibinfo {volume} {15}},\ \bibinfo {pages} {503} (\bibinfo {year}
  {2019})}\BibitemShut {NoStop}%
\bibitem [{\citenamefont {Gottesman}\ \emph {et~al.}(2001)\citenamefont
  {Gottesman}, \citenamefont {Kitaev},\ and\ \citenamefont
  {Preskill}}]{GKP2001}%
  \BibitemOpen
  \bibfield  {author} {\bibinfo {author} {\bibfnamefont {D.}~\bibnamefont
  {Gottesman}}, \bibinfo {author} {\bibfnamefont {A.}~\bibnamefont {Kitaev}},\
  and\ \bibinfo {author} {\bibfnamefont {J.}~\bibnamefont {Preskill}},\
  }\bibfield  {title} {\bibinfo {title} {Encoding a qubit in an oscillator},\
  }\href {https://doi.org/10.1103/PhysRevA.64.012310} {\bibfield  {journal}
  {\bibinfo  {journal} {Phys. Rev. A}\ }\textbf {\bibinfo {volume} {64}},\
  \bibinfo {pages} {012310} (\bibinfo {year} {2001})}\BibitemShut {NoStop}%
\bibitem [{\citenamefont {Tosta}\ \emph {et~al.}(2022)\citenamefont {Tosta},
  \citenamefont {Maciel},\ and\ \citenamefont {Aolita}}]{tosta_grand_2022}%
  \BibitemOpen
  \bibfield  {author} {\bibinfo {author} {\bibfnamefont {A.~D.~C.}\
  \bibnamefont {Tosta}}, \bibinfo {author} {\bibfnamefont {T.~O.}\ \bibnamefont
  {Maciel}},\ and\ \bibinfo {author} {\bibfnamefont {L.}~\bibnamefont
  {Aolita}},\ }\href {https://doi.org/10.48550/arXiv.2206.01751} {\bibinfo
  {title} {Grand {Unification} of continuous-variable codes}} (\bibinfo {year}
  {2022}),\ \bibinfo {note} {arXiv:2206.01751 [quant-ph]}\BibitemShut {NoStop}%
\bibitem [{\citenamefont {Campagne-Ibarcq}\ \emph {et~al.}(2020)\citenamefont
  {Campagne-Ibarcq}, \citenamefont {Eickbusch}, \citenamefont {Touzard},
  \citenamefont {Zalys-Geller}, \citenamefont {Frattini}, \citenamefont
  {Sivak}, \citenamefont {Reinhold}, \citenamefont {Puri}, \citenamefont
  {Shankar}, \citenamefont {Schoelkopf}, \citenamefont {Frunzio}, \citenamefont
  {Mirrahimi},\ and\ \citenamefont {Devoret}}]{Campagne-Ibarcq2020}%
  \BibitemOpen
  \bibfield  {author} {\bibinfo {author} {\bibfnamefont {P.}~\bibnamefont
  {Campagne-Ibarcq}}, \bibinfo {author} {\bibfnamefont {A.}~\bibnamefont
  {Eickbusch}}, \bibinfo {author} {\bibfnamefont {S.}~\bibnamefont {Touzard}},
  \bibinfo {author} {\bibfnamefont {E.}~\bibnamefont {Zalys-Geller}}, \bibinfo
  {author} {\bibfnamefont {N.~E.}\ \bibnamefont {Frattini}}, \bibinfo {author}
  {\bibfnamefont {V.~V.}\ \bibnamefont {Sivak}}, \bibinfo {author}
  {\bibfnamefont {P.}~\bibnamefont {Reinhold}}, \bibinfo {author}
  {\bibfnamefont {S.}~\bibnamefont {Puri}}, \bibinfo {author} {\bibfnamefont
  {S.}~\bibnamefont {Shankar}}, \bibinfo {author} {\bibfnamefont {R.~J.}\
  \bibnamefont {Schoelkopf}}, \bibinfo {author} {\bibfnamefont
  {L.}~\bibnamefont {Frunzio}}, \bibinfo {author} {\bibfnamefont
  {M.}~\bibnamefont {Mirrahimi}},\ and\ \bibinfo {author} {\bibfnamefont
  {M.~H.}\ \bibnamefont {Devoret}},\ }\bibfield  {title} {\bibinfo {title}
  {Quantum error correction of a qubit encoded in grid states of an
  oscillator},\ }\href {https://doi.org/10.1038/s41586-020-2603-3} {\bibfield
  {journal} {\bibinfo  {journal} {Nature}\ }\textbf {\bibinfo {volume} {584}},\
  \bibinfo {pages} {368} (\bibinfo {year} {2020})}\BibitemShut {NoStop}%
\bibitem [{\citenamefont {Sivak}\ \emph {et~al.}(2023)\citenamefont {Sivak},
  \citenamefont {Eickbusch}, \citenamefont {Royer}, \citenamefont {Singh},
  \citenamefont {Tsioutsios}, \citenamefont {Ganjam}, \citenamefont {Miano},
  \citenamefont {Brock}, \citenamefont {Ding}, \citenamefont {Frunzio},
  \citenamefont {Girvin}, \citenamefont {Schoelkopf},\ and\ \citenamefont
  {Devoret}}]{Sivak2023}%
  \BibitemOpen
  \bibfield  {author} {\bibinfo {author} {\bibfnamefont {V.~V.}\ \bibnamefont
  {Sivak}}, \bibinfo {author} {\bibfnamefont {A.}~\bibnamefont {Eickbusch}},
  \bibinfo {author} {\bibfnamefont {B.}~\bibnamefont {Royer}}, \bibinfo
  {author} {\bibfnamefont {S.}~\bibnamefont {Singh}}, \bibinfo {author}
  {\bibfnamefont {I.}~\bibnamefont {Tsioutsios}}, \bibinfo {author}
  {\bibfnamefont {S.}~\bibnamefont {Ganjam}}, \bibinfo {author} {\bibfnamefont
  {A.}~\bibnamefont {Miano}}, \bibinfo {author} {\bibfnamefont {B.~L.}\
  \bibnamefont {Brock}}, \bibinfo {author} {\bibfnamefont {A.~Z.}\ \bibnamefont
  {Ding}}, \bibinfo {author} {\bibfnamefont {L.}~\bibnamefont {Frunzio}},
  \bibinfo {author} {\bibfnamefont {S.~M.}\ \bibnamefont {Girvin}}, \bibinfo
  {author} {\bibfnamefont {R.~J.}\ \bibnamefont {Schoelkopf}},\ and\ \bibinfo
  {author} {\bibfnamefont {M.~H.}\ \bibnamefont {Devoret}},\ }\bibfield
  {title} {\bibinfo {title} {Real-time quantum error correction beyond
  break-even},\ }\href {https://doi.org/10.1038/s41586-023-05782-6} {\bibfield
  {journal} {\bibinfo  {journal} {Nature}\ }\textbf {\bibinfo {volume} {616}},\
  \bibinfo {pages} {50} (\bibinfo {year} {2023})}\BibitemShut {NoStop}%
\bibitem [{\citenamefont {Cahill}\ and\ \citenamefont
  {Glauber}(1969)}]{CahillGlauber69}%
  \BibitemOpen
  \bibfield  {author} {\bibinfo {author} {\bibfnamefont {K.~E.}\ \bibnamefont
  {Cahill}}\ and\ \bibinfo {author} {\bibfnamefont {R.~J.}\ \bibnamefont
  {Glauber}},\ }\bibfield  {title} {\bibinfo {title} {Ordered expansions in
  boson amplitude operators},\ }\href
  {https://doi.org/10.1103/PhysRev.177.1857} {\bibfield  {journal} {\bibinfo
  {journal} {Phys. Rev.}\ }\textbf {\bibinfo {volume} {177}},\ \bibinfo {pages}
  {1857} (\bibinfo {year} {1969})}\BibitemShut {NoStop}%
\bibitem [{\citenamefont {Krastanov}\ \emph {et~al.}(2015)\citenamefont
  {Krastanov}, \citenamefont {Albert}, \citenamefont {Shen}, \citenamefont
  {Zou}, \citenamefont {Heeres}, \citenamefont {Vlastakis}, \citenamefont
  {Schoelkopf},\ and\ \citenamefont {Jiang}}]{SNAPgateTHY2015}%
  \BibitemOpen
  \bibfield  {author} {\bibinfo {author} {\bibfnamefont {S.}~\bibnamefont
  {Krastanov}}, \bibinfo {author} {\bibfnamefont {V.~V.}\ \bibnamefont
  {Albert}}, \bibinfo {author} {\bibfnamefont {C.}~\bibnamefont {Shen}},
  \bibinfo {author} {\bibfnamefont {C.-L.}\ \bibnamefont {Zou}}, \bibinfo
  {author} {\bibfnamefont {R.~W.}\ \bibnamefont {Heeres}}, \bibinfo {author}
  {\bibfnamefont {B.}~\bibnamefont {Vlastakis}}, \bibinfo {author}
  {\bibfnamefont {R.~J.}\ \bibnamefont {Schoelkopf}},\ and\ \bibinfo {author}
  {\bibfnamefont {L.}~\bibnamefont {Jiang}},\ }\bibfield  {title} {\bibinfo
  {title} {Universal control of an oscillator with dispersive coupling to a
  qubit},\ }\href {https://doi.org/10.1103/PhysRevA.92.040303} {\bibfield
  {journal} {\bibinfo  {journal} {Phys. Rev. A}\ }\textbf {\bibinfo {volume}
  {92}},\ \bibinfo {pages} {040303} (\bibinfo {year} {2015})}\BibitemShut
  {NoStop}%
\bibitem [{\citenamefont {Heeres}\ \emph {et~al.}(2015)\citenamefont {Heeres},
  \citenamefont {Vlastakis}, \citenamefont {Holland}, \citenamefont
  {Krastanov}, \citenamefont {Albert}, \citenamefont {Frunzio}, \citenamefont
  {Jiang},\ and\ \citenamefont {Schoelkopf}}]{SNAPgateEXP2015}%
  \BibitemOpen
  \bibfield  {author} {\bibinfo {author} {\bibfnamefont {R.~W.}\ \bibnamefont
  {Heeres}}, \bibinfo {author} {\bibfnamefont {B.}~\bibnamefont {Vlastakis}},
  \bibinfo {author} {\bibfnamefont {E.}~\bibnamefont {Holland}}, \bibinfo
  {author} {\bibfnamefont {S.}~\bibnamefont {Krastanov}}, \bibinfo {author}
  {\bibfnamefont {V.~V.}\ \bibnamefont {Albert}}, \bibinfo {author}
  {\bibfnamefont {L.}~\bibnamefont {Frunzio}}, \bibinfo {author} {\bibfnamefont
  {L.}~\bibnamefont {Jiang}},\ and\ \bibinfo {author} {\bibfnamefont {R.~J.}\
  \bibnamefont {Schoelkopf}},\ }\bibfield  {title} {\bibinfo {title} {Cavity
  state manipulation using photon-number selective phase gates},\ }\href
  {https://doi.org/10.1103/PhysRevLett.115.137002} {\bibfield  {journal}
  {\bibinfo  {journal} {Phys. Rev. Lett.}\ }\textbf {\bibinfo {volume} {115}},\
  \bibinfo {pages} {137002} (\bibinfo {year} {2015})}\BibitemShut {NoStop}%
\bibitem [{\citenamefont {Michael}\ \emph {et~al.}(2016)\citenamefont
  {Michael}, \citenamefont {Silveri}, \citenamefont {Brierley}, \citenamefont
  {Albert}, \citenamefont {Salmilehto}, \citenamefont {Jiang},\ and\
  \citenamefont {Girvin}}]{Michael16}%
  \BibitemOpen
  \bibfield  {author} {\bibinfo {author} {\bibfnamefont {M.~H.}\ \bibnamefont
  {Michael}}, \bibinfo {author} {\bibfnamefont {M.}~\bibnamefont {Silveri}},
  \bibinfo {author} {\bibfnamefont {R.~T.}\ \bibnamefont {Brierley}}, \bibinfo
  {author} {\bibfnamefont {V.~V.}\ \bibnamefont {Albert}}, \bibinfo {author}
  {\bibfnamefont {J.}~\bibnamefont {Salmilehto}}, \bibinfo {author}
  {\bibfnamefont {L.}~\bibnamefont {Jiang}},\ and\ \bibinfo {author}
  {\bibfnamefont {S.~M.}\ \bibnamefont {Girvin}},\ }\bibfield  {title}
  {\bibinfo {title} {New class of quantum error-correcting codes for a bosonic
  mode},\ }\href {https://doi.org/10.1103/PhysRevX.6.031006} {\bibfield
  {journal} {\bibinfo  {journal} {Phys. Rev. X}\ }\textbf {\bibinfo {volume}
  {6}},\ \bibinfo {pages} {031006} (\bibinfo {year} {2016})}\BibitemShut
  {NoStop}%
\bibitem [{\citenamefont {Albert}\ \emph {et~al.}(2018)\citenamefont {Albert},
  \citenamefont {Noh}, \citenamefont {Duivenvoorden}, \citenamefont {Young},
  \citenamefont {Brierley}, \citenamefont {Reinhold}, \citenamefont {Vuillot},
  \citenamefont {Li}, \citenamefont {Shen}, \citenamefont {Girvin},
  \citenamefont {Terhal},\ and\ \citenamefont {Jiang}}]{albert2018}%
  \BibitemOpen
  \bibfield  {author} {\bibinfo {author} {\bibfnamefont {V.~V.}\ \bibnamefont
  {Albert}}, \bibinfo {author} {\bibfnamefont {K.}~\bibnamefont {Noh}},
  \bibinfo {author} {\bibfnamefont {K.}~\bibnamefont {Duivenvoorden}}, \bibinfo
  {author} {\bibfnamefont {D.~J.}\ \bibnamefont {Young}}, \bibinfo {author}
  {\bibfnamefont {R.~T.}\ \bibnamefont {Brierley}}, \bibinfo {author}
  {\bibfnamefont {P.}~\bibnamefont {Reinhold}}, \bibinfo {author}
  {\bibfnamefont {C.}~\bibnamefont {Vuillot}}, \bibinfo {author} {\bibfnamefont
  {L.}~\bibnamefont {Li}}, \bibinfo {author} {\bibfnamefont {C.}~\bibnamefont
  {Shen}}, \bibinfo {author} {\bibfnamefont {S.~M.}\ \bibnamefont {Girvin}},
  \bibinfo {author} {\bibfnamefont {B.~M.}\ \bibnamefont {Terhal}},\ and\
  \bibinfo {author} {\bibfnamefont {L.}~\bibnamefont {Jiang}},\ }\bibfield
  {title} {\bibinfo {title} {Performance and structure of single-mode bosonic
  codes},\ }\href {https://doi.org/10.1103/PhysRevA.97.032346} {\bibfield
  {journal} {\bibinfo  {journal} {Phys. Rev. A}\ }\textbf {\bibinfo {volume}
  {97}},\ \bibinfo {pages} {032346} (\bibinfo {year} {2018})}\BibitemShut
  {NoStop}%
\bibitem [{\citenamefont {Ouyang}\ and\ \citenamefont
  {Campbell}(2021)}]{Yingkai2021}%
  \BibitemOpen
  \bibfield  {author} {\bibinfo {author} {\bibfnamefont {Y.}~\bibnamefont
  {Ouyang}}\ and\ \bibinfo {author} {\bibfnamefont {E.~T.}\ \bibnamefont
  {Campbell}},\ }\bibfield  {title} {\bibinfo {title} {Trade-offs on number and
  phase shift resilience in bosonic quantum codes},\ }\href
  {https://doi.org/10.1109/TIT.2021.3102873} {\bibfield  {journal} {\bibinfo
  {journal} {IEEE Transactions on Information Theory}\ ,\ \bibinfo {pages} {1}}
  (\bibinfo {year} {2021})}\BibitemShut {NoStop}%
\bibitem [{\citenamefont {Cochrane}\ \emph {et~al.}(1999)\citenamefont
  {Cochrane}, \citenamefont {Milburn},\ and\ \citenamefont
  {Munro}}]{Cochrane99}%
  \BibitemOpen
  \bibfield  {author} {\bibinfo {author} {\bibfnamefont {P.~T.}\ \bibnamefont
  {Cochrane}}, \bibinfo {author} {\bibfnamefont {G.~J.}\ \bibnamefont
  {Milburn}},\ and\ \bibinfo {author} {\bibfnamefont {W.~J.}\ \bibnamefont
  {Munro}},\ }\bibfield  {title} {\bibinfo {title} {Macroscopically distinct
  quantum-superposition states as a bosonic code for amplitude damping},\
  }\href {https://doi.org/10.1103/PhysRevA.59.2631} {\bibfield  {journal}
  {\bibinfo  {journal} {Phys. Rev. A}\ }\textbf {\bibinfo {volume} {59}},\
  \bibinfo {pages} {2631} (\bibinfo {year} {1999})}\BibitemShut {NoStop}%
\bibitem [{\citenamefont {Ralph}\ \emph {et~al.}(2003)\citenamefont {Ralph},
  \citenamefont {Gilchrist}, \citenamefont {Milburn}, \citenamefont {Munro},\
  and\ \citenamefont {Glancy}}]{Ralph2003}%
  \BibitemOpen
  \bibfield  {author} {\bibinfo {author} {\bibfnamefont {T.~C.}\ \bibnamefont
  {Ralph}}, \bibinfo {author} {\bibfnamefont {A.}~\bibnamefont {Gilchrist}},
  \bibinfo {author} {\bibfnamefont {G.~J.}\ \bibnamefont {Milburn}}, \bibinfo
  {author} {\bibfnamefont {W.~J.}\ \bibnamefont {Munro}},\ and\ \bibinfo
  {author} {\bibfnamefont {S.}~\bibnamefont {Glancy}},\ }\bibfield  {title}
  {\bibinfo {title} {Quantum computation with optical coherent states},\ }\href
  {https://doi.org/10.1103/PhysRevA.68.042319} {\bibfield  {journal} {\bibinfo
  {journal} {Phys. Rev. A}\ }\textbf {\bibinfo {volume} {68}},\ \bibinfo
  {pages} {042319} (\bibinfo {year} {2003})}\BibitemShut {NoStop}%
\bibitem [{\citenamefont {Leghtas}\ \emph {et~al.}(2013)\citenamefont
  {Leghtas}, \citenamefont {Kirchmair}, \citenamefont {Vlastakis},
  \citenamefont {Schoelkopf}, \citenamefont {Devoret},\ and\ \citenamefont
  {Mirrahimi}}]{Zaki}%
  \BibitemOpen
  \bibfield  {author} {\bibinfo {author} {\bibfnamefont {Z.}~\bibnamefont
  {Leghtas}}, \bibinfo {author} {\bibfnamefont {G.}~\bibnamefont {Kirchmair}},
  \bibinfo {author} {\bibfnamefont {B.}~\bibnamefont {Vlastakis}}, \bibinfo
  {author} {\bibfnamefont {R.~J.}\ \bibnamefont {Schoelkopf}}, \bibinfo
  {author} {\bibfnamefont {M.~H.}\ \bibnamefont {Devoret}},\ and\ \bibinfo
  {author} {\bibfnamefont {M.}~\bibnamefont {Mirrahimi}},\ }\bibfield  {title}
  {\bibinfo {title} {Hardware-efficient autonomous quantum memory protection},\
  }\href {https://doi.org/10.1103/PhysRevLett.111.120501} {\bibfield  {journal}
  {\bibinfo  {journal} {Phys. Rev. Lett.}\ }\textbf {\bibinfo {volume} {111}},\
  \bibinfo {pages} {120501} (\bibinfo {year} {2013})}\BibitemShut {NoStop}%
\bibitem [{\citenamefont {Mirrahimi}\ \emph {et~al.}(2014)\citenamefont
  {Mirrahimi}, \citenamefont {Leghtas}, \citenamefont {Albert}, \citenamefont
  {Touzard}, \citenamefont {Schoelkopf}, \citenamefont {Jiang},\ and\
  \citenamefont {Devoret}}]{Mirrahimi_2014}%
  \BibitemOpen
  \bibfield  {author} {\bibinfo {author} {\bibfnamefont {M.}~\bibnamefont
  {Mirrahimi}}, \bibinfo {author} {\bibfnamefont {Z.}~\bibnamefont {Leghtas}},
  \bibinfo {author} {\bibfnamefont {V.~V.}\ \bibnamefont {Albert}}, \bibinfo
  {author} {\bibfnamefont {S.}~\bibnamefont {Touzard}}, \bibinfo {author}
  {\bibfnamefont {R.~J.}\ \bibnamefont {Schoelkopf}}, \bibinfo {author}
  {\bibfnamefont {L.}~\bibnamefont {Jiang}},\ and\ \bibinfo {author}
  {\bibfnamefont {M.~H.}\ \bibnamefont {Devoret}},\ }\bibfield  {title}
  {\bibinfo {title} {Dynamically protected cat-qubits: a new paradigm for
  universal quantum computation},\ }\href
  {https://doi.org/10.1088/1367-2630/16/4/045014} {\bibfield  {journal}
  {\bibinfo  {journal} {New Journal of Physics}\ }\textbf {\bibinfo {volume}
  {16}},\ \bibinfo {pages} {045014} (\bibinfo {year} {2014})}\BibitemShut
  {NoStop}%
\bibitem [{\citenamefont {Carruthers}\ and\ \citenamefont
  {Nieto}(1968)}]{CarrNietRMP68}%
  \BibitemOpen
  \bibfield  {author} {\bibinfo {author} {\bibfnamefont {P.}~\bibnamefont
  {Carruthers}}\ and\ \bibinfo {author} {\bibfnamefont {M.~M.}\ \bibnamefont
  {Nieto}},\ }\bibfield  {title} {\bibinfo {title} {Phase and angle variables
  in quantum mechanics},\ }\href {https://dx.doi.org/10.1103/RevModPhys.40.411}
  {\bibfield  {journal} {\bibinfo  {journal} {Rev. Mod. Phys.}\ }\textbf
  {\bibinfo {volume} {40}},\ \bibinfo {pages} {411} (\bibinfo {year}
  {1968})}\BibitemShut {NoStop}%
\bibitem [{\citenamefont {Susskind}\ and\ \citenamefont
  {Glogower}(1964)}]{Suss64}%
  \BibitemOpen
  \bibfield  {author} {\bibinfo {author} {\bibfnamefont {L.}~\bibnamefont
  {Susskind}}\ and\ \bibinfo {author} {\bibfnamefont {J.}~\bibnamefont
  {Glogower}},\ }\bibfield  {title} {\bibinfo {title} {Quantum mechanical phase
  and time operator},\ }\href
  {https://doi.org/10.1103/PhysicsPhysiqueFizika.1.49} {\bibfield  {journal}
  {\bibinfo  {journal} {Physics}\ }\textbf {\bibinfo {volume} {1}},\ \bibinfo
  {pages} {49} (\bibinfo {year} {1964})}\BibitemShut {NoStop}%
\bibitem [{\citenamefont {Combes}\ \emph {et~al.}(2024)\citenamefont {Combes},
  \citenamefont {Albert}, \citenamefont {Noh}, \citenamefont {Woods},
  \citenamefont {Grimsmo},\ and\ \citenamefont
  {Baragiola}}]{CombesAlbertetal2021}%
  \BibitemOpen
  \bibfield  {author} {\bibinfo {author} {\bibfnamefont {J.}~\bibnamefont
  {Combes}}, \bibinfo {author} {\bibfnamefont {V.}~\bibnamefont {Albert}},
  \bibinfo {author} {\bibfnamefont {K.}~\bibnamefont {Noh}}, \bibinfo {author}
  {\bibfnamefont {M.}~\bibnamefont {Woods}}, \bibinfo {author} {\bibfnamefont
  {A.}~\bibnamefont {Grimsmo}},\ and\ \bibinfo {author} {\bibfnamefont
  {B.}~\bibnamefont {Baragiola}},\ }\bibfield  {title} {\bibinfo {title}
  {Modular phase: measurement, metrology, capacities, and codes},\ }\href@noop
  {} {\bibfield  {journal} {\bibinfo  {journal} {In preparation}\ } (\bibinfo
  {year} {2024})}\BibitemShut {NoStop}%
\bibitem [{\citenamefont {Baragiola}()}]{BaraG}%
  \BibitemOpen
  \bibfield  {author} {\bibinfo {author} {\bibfnamefont {B.~Q.}\ \bibnamefont
  {Baragiola}},\ }\href@noop {} {}\bibinfo {howpublished} {personal
  communication}\BibitemShut {NoStop}%
\bibitem [{\citenamefont {Garc\'{\i}a-\'Alvarez}\ \emph
  {et~al.}(2020)\citenamefont {Garc\'{\i}a-\'Alvarez}, \citenamefont
  {Calcluth}, \citenamefont {Ferraro},\ and\ \citenamefont
  {Ferrini}}]{AlvCalcFerrFerr20}%
  \BibitemOpen
  \bibfield  {author} {\bibinfo {author} {\bibfnamefont {L.}~\bibnamefont
  {Garc\'{\i}a-\'Alvarez}}, \bibinfo {author} {\bibfnamefont {C.}~\bibnamefont
  {Calcluth}}, \bibinfo {author} {\bibfnamefont {A.}~\bibnamefont {Ferraro}},\
  and\ \bibinfo {author} {\bibfnamefont {G.}~\bibnamefont {Ferrini}},\
  }\bibfield  {title} {\bibinfo {title} {Efficient simulatability of
  continuous-variable circuits with large wigner negativity},\ }\href
  {https://doi.org/10.1103/PhysRevResearch.2.043322} {\bibfield  {journal}
  {\bibinfo  {journal} {Phys. Rev. Research}\ }\textbf {\bibinfo {volume}
  {2}},\ \bibinfo {pages} {043322} (\bibinfo {year} {2020})}\BibitemShut
  {NoStop}%
\bibitem [{\citenamefont {Gertler}\ \emph {et~al.}(2021)\citenamefont
  {Gertler}, \citenamefont {Baker}, \citenamefont {Li}, \citenamefont {Shirol},
  \citenamefont {Koch},\ and\ \citenamefont {Wang}}]{Gertler21}%
  \BibitemOpen
  \bibfield  {author} {\bibinfo {author} {\bibfnamefont {J.~M.}\ \bibnamefont
  {Gertler}}, \bibinfo {author} {\bibfnamefont {B.}~\bibnamefont {Baker}},
  \bibinfo {author} {\bibfnamefont {J.}~\bibnamefont {Li}}, \bibinfo {author}
  {\bibfnamefont {S.}~\bibnamefont {Shirol}}, \bibinfo {author} {\bibfnamefont
  {J.}~\bibnamefont {Koch}},\ and\ \bibinfo {author} {\bibfnamefont
  {C.}~\bibnamefont {Wang}},\ }\bibfield  {title} {\bibinfo {title} {Protecting
  a bosonic qubit with autonomous quantum error correction},\ }\href
  {https://doi.org/10.1038/s41586-021-03257-0} {\bibfield  {journal} {\bibinfo
  {journal} {Nature}\ }\textbf {\bibinfo {volume} {590}},\ \bibinfo {pages}
  {243} (\bibinfo {year} {2021})}\BibitemShut {NoStop}%
\bibitem [{\citenamefont {Knill}(2005)}]{Knill2005}%
  \BibitemOpen
  \bibfield  {author} {\bibinfo {author} {\bibfnamefont {E.}~\bibnamefont
  {Knill}},\ }\bibfield  {title} {\bibinfo {title} {Quantum computing with
  realistically noisy devices},\ }\href {https://doi.org/10.1038/nature03350}
  {\bibfield  {journal} {\bibinfo  {journal} {Nature}\ }\textbf {\bibinfo
  {volume} {434}},\ \bibinfo {pages} {39} (\bibinfo {year} {2005})}\BibitemShut
  {NoStop}%
\bibitem [{\citenamefont {Leung}\ \emph {et~al.}(1997)\citenamefont {Leung},
  \citenamefont {Nielsen}, \citenamefont {Chuang},\ and\ \citenamefont
  {Yamamoto}}]{ApproxLeung97}%
  \BibitemOpen
  \bibfield  {author} {\bibinfo {author} {\bibfnamefont {D.~W.}\ \bibnamefont
  {Leung}}, \bibinfo {author} {\bibfnamefont {M.~A.}\ \bibnamefont {Nielsen}},
  \bibinfo {author} {\bibfnamefont {I.~L.}\ \bibnamefont {Chuang}},\ and\
  \bibinfo {author} {\bibfnamefont {Y.}~\bibnamefont {Yamamoto}},\ }\bibfield
  {title} {\bibinfo {title} {Approximate quantum error correction can lead to
  better codes},\ }\href {https://doi.org/10.1103/PhysRevA.56.2567} {\bibfield
  {journal} {\bibinfo  {journal} {Phys. Rev. A}\ }\textbf {\bibinfo {volume}
  {56}},\ \bibinfo {pages} {2567} (\bibinfo {year} {1997})}\BibitemShut
  {NoStop}%
\bibitem [{\citenamefont {Knill}\ and\ \citenamefont
  {Laflamme}(1997)}]{KnillLaflamme97}%
  \BibitemOpen
  \bibfield  {author} {\bibinfo {author} {\bibfnamefont {E.}~\bibnamefont
  {Knill}}\ and\ \bibinfo {author} {\bibfnamefont {R.}~\bibnamefont
  {Laflamme}},\ }\bibfield  {title} {\bibinfo {title} {Theory of quantum
  error-correcting codes},\ }\href {https://doi.org/10.1103/PhysRevA.55.900}
  {\bibfield  {journal} {\bibinfo  {journal} {Phys. Rev. A}\ }\textbf {\bibinfo
  {volume} {55}},\ \bibinfo {pages} {900} (\bibinfo {year} {1997})}\BibitemShut
  {NoStop}%
\bibitem [{\citenamefont {Li}\ \emph {et~al.}(2021)\citenamefont {Li},
  \citenamefont {Young}, \citenamefont {Albert}, \citenamefont {Noh},
  \citenamefont {Zou},\ and\ \citenamefont {Jiang}}]{PhaseEngcodes}%
  \BibitemOpen
  \bibfield  {author} {\bibinfo {author} {\bibfnamefont {L.}~\bibnamefont
  {Li}}, \bibinfo {author} {\bibfnamefont {D.~J.}\ \bibnamefont {Young}},
  \bibinfo {author} {\bibfnamefont {V.~V.}\ \bibnamefont {Albert}}, \bibinfo
  {author} {\bibfnamefont {K.}~\bibnamefont {Noh}}, \bibinfo {author}
  {\bibfnamefont {C.-L.}\ \bibnamefont {Zou}},\ and\ \bibinfo {author}
  {\bibfnamefont {L.}~\bibnamefont {Jiang}},\ }\bibfield  {title} {\bibinfo
  {title} {Phase-engineered bosonic quantum codes},\ }\href
  {https://doi.org/10.1103/PhysRevA.103.062427} {\bibfield  {journal} {\bibinfo
   {journal} {Phys. Rev. A}\ }\textbf {\bibinfo {volume} {103}},\ \bibinfo
  {pages} {062427} (\bibinfo {year} {2021})}\BibitemShut {NoStop}%
\bibitem [{\citenamefont {Gao}\ \emph {et~al.}(2019)\citenamefont {Gao},
  \citenamefont {Lester}, \citenamefont {Chou}, \citenamefont {Frunzio},
  \citenamefont {Devoret}, \citenamefont {Jiang}, \citenamefont {Girvin},\ and\
  \citenamefont {Schoelkopf}}]{GaoLesCho2019}%
  \BibitemOpen
  \bibfield  {author} {\bibinfo {author} {\bibfnamefont {Y.~Y.}\ \bibnamefont
  {Gao}}, \bibinfo {author} {\bibfnamefont {B.~J.}\ \bibnamefont {Lester}},
  \bibinfo {author} {\bibfnamefont {K.~S.}\ \bibnamefont {Chou}}, \bibinfo
  {author} {\bibfnamefont {L.}~\bibnamefont {Frunzio}}, \bibinfo {author}
  {\bibfnamefont {M.~H.}\ \bibnamefont {Devoret}}, \bibinfo {author}
  {\bibfnamefont {L.}~\bibnamefont {Jiang}}, \bibinfo {author} {\bibfnamefont
  {S.~M.}\ \bibnamefont {Girvin}},\ and\ \bibinfo {author} {\bibfnamefont
  {R.~J.}\ \bibnamefont {Schoelkopf}},\ }\bibfield  {title} {\bibinfo {title}
  {Entanglement of bosonic modes through an engineered exchange interaction},\
  }\href {https://doi.org/10.1038/s41586-019-0970-4} {\bibfield  {journal}
  {\bibinfo  {journal} {Nature}\ }\textbf {\bibinfo {volume} {566}},\ \bibinfo
  {pages} {509} (\bibinfo {year} {2019})}\BibitemShut {NoStop}%
\bibitem [{\citenamefont {Chamberland}\ \emph {et~al.}(2022)\citenamefont
  {Chamberland}, \citenamefont {Noh}, \citenamefont {Arrangoiz-Arriola},
  \citenamefont {Campbell}, \citenamefont {Hann}, \citenamefont {Iverson},
  \citenamefont {Putterman}, \citenamefont {Bohdanowicz}, \citenamefont
  {Flammia}, \citenamefont {Keller}, \citenamefont {Refael}, \citenamefont
  {Preskill}, \citenamefont {Jiang}, \citenamefont {Safavi-Naeini},
  \citenamefont {Painter},\ and\ \citenamefont {Brand\~ao}}]{Chamberland_2022}%
  \BibitemOpen
  \bibfield  {author} {\bibinfo {author} {\bibfnamefont {C.}~\bibnamefont
  {Chamberland}}, \bibinfo {author} {\bibfnamefont {K.}~\bibnamefont {Noh}},
  \bibinfo {author} {\bibfnamefont {P.}~\bibnamefont {Arrangoiz-Arriola}},
  \bibinfo {author} {\bibfnamefont {E.~T.}\ \bibnamefont {Campbell}}, \bibinfo
  {author} {\bibfnamefont {C.~T.}\ \bibnamefont {Hann}}, \bibinfo {author}
  {\bibfnamefont {J.}~\bibnamefont {Iverson}}, \bibinfo {author} {\bibfnamefont
  {H.}~\bibnamefont {Putterman}}, \bibinfo {author} {\bibfnamefont {T.~C.}\
  \bibnamefont {Bohdanowicz}}, \bibinfo {author} {\bibfnamefont {S.~T.}\
  \bibnamefont {Flammia}}, \bibinfo {author} {\bibfnamefont {A.}~\bibnamefont
  {Keller}}, \bibinfo {author} {\bibfnamefont {G.}~\bibnamefont {Refael}},
  \bibinfo {author} {\bibfnamefont {J.}~\bibnamefont {Preskill}}, \bibinfo
  {author} {\bibfnamefont {L.}~\bibnamefont {Jiang}}, \bibinfo {author}
  {\bibfnamefont {A.~H.}\ \bibnamefont {Safavi-Naeini}}, \bibinfo {author}
  {\bibfnamefont {O.}~\bibnamefont {Painter}},\ and\ \bibinfo {author}
  {\bibfnamefont {F.~G.}\ \bibnamefont {Brand\~ao}},\ }\bibfield  {title}
  {\bibinfo {title} {Building a fault-tolerant quantum computer using
  concatenated cat codes},\ }\href
  {https://doi.org/10.1103/PRXQuantum.3.010329} {\bibfield  {journal} {\bibinfo
   {journal} {PRX Quantum}\ }\textbf {\bibinfo {volume} {3}},\ \bibinfo {pages}
  {010329} (\bibinfo {year} {2022})}\BibitemShut {NoStop}%
\bibitem [{\citenamefont {Gautier}\ \emph {et~al.}(2023)\citenamefont
  {Gautier}, \citenamefont {Mirrahimi},\ and\ \citenamefont
  {Sarlette}}]{Gautier_2023}%
  \BibitemOpen
  \bibfield  {author} {\bibinfo {author} {\bibfnamefont {R.}~\bibnamefont
  {Gautier}}, \bibinfo {author} {\bibfnamefont {M.}~\bibnamefont {Mirrahimi}},\
  and\ \bibinfo {author} {\bibfnamefont {A.}~\bibnamefont {Sarlette}},\
  }\bibfield  {title} {\bibinfo {title} {Designing high-fidelity zeno gates for
  dissipative cat qubits},\ }\href
  {https://doi.org/10.1103/PRXQuantum.4.040316} {\bibfield  {journal} {\bibinfo
   {journal} {PRX Quantum}\ }\textbf {\bibinfo {volume} {4}},\ \bibinfo {pages}
  {040316} (\bibinfo {year} {2023})}\BibitemShut {NoStop}%
\end{thebibliography}%
%============================================
\vspace{2em}

%============================================
% Appendix
%============================================
\newpage\newpage
%\onecolumngrid
\appendix

%============================================
\section{Derivation of \cref{eq:general_function_gate_prop,eq:linear_error_modification}} \label{sec:derivations_appendix}
%============================================
Beginning with \cref{eq:errprop}, we take $\hat G =e^{i f(\hat{n})}$ and insert the identity $\hat G\dg \hat G$, on the RHS 
\begin{equation}
    e^{i f(\hat{n})} \E{k}{\theta} = e^{i f(\hat{n})} \E{k}{\theta} e^{-i f(\hat{n})} e^{i f(\hat{n})}.
\end{equation}
For $k < 0$ this gives
\begin{align}
\begin{aligned}
    e^{i f(\hat{n})} \E{k}{\theta} &= \left( e^{i f(\hat{n})} e^{i\theta \hat{n}} \sum_{n=0}^{\infty} \op{n}{n+|k|} e^{-i f(\hat{n})} \right) e^{i f(\hat{n})} \\
    &= \left(\sum_{n=0}^{\infty} e^{if(n)}e^{-if(n+|k|)}e^{i\theta n} \op{n}{n+|k|} \right) e^{i f(\hat{n})} \\
    &= \exp \left[ if(\hat{n}) -if(\hat{n} + |k|\hat{I}) \right] \E{k}{\theta} e^{i f(\hat{n})}.
\end{aligned}
\end{align}
Similarly for $k \ge 0$, we get
\begin{align}
\begin{aligned}
    e^{i f(\hat{n})} \E{k}{\theta} &= \left( e^{i f(\hat{n})} \sum_{n=0}^{\infty} \op{n+k}{n} e^{i\theta \hat{n}} e^{-i f(\hat{n})} \right) e^{i f(\hat{n})} \\
   &= \left(\sum_{n=0}^{\infty} e^{if(n+k)}e^{-if(n)} \op{n+k}{n} e^{i\theta n} \right) e^{i f(\hat{n})} \\
  &  = \exp \left[ if(\hat{n}) -if(\hat{n} - k\hat{I}) \right] \E{k}{\theta} e^{i f(\hat{n})}.
\end{aligned}
\end{align}
These are both of the desired form, thus \cref{eq:general_function_gate_prop} holds for all $k$.

To derive \cref{eq:linear_error_modification}, we again treat the $k<0$ and $k \ge 0$ cases separately. For $k<0$,
\begin{equation}
    e^{i\phi \hat{n}} \E{k}{\theta} = e^{i\phi \hat{n}}e^{i\theta \hat{n}} \sum_{n=0}^{\infty} \op{n}{n+|k|} = \E{k}{\theta + \phi}
\end{equation}
trivially. For $k \ge 0$,
\begin{align}
\begin{aligned}
    e^{i\phi \hat{n}} \E{k}{\theta} &= e^{i\phi \hat{n}} \sum_{n=0}^{\infty} \op{n+k}{n} e^{i\theta \hat{n}} \\
    &= e^{i\phi k} \sum_{n=0}^{\infty} \op{n+k}{n} e^{i\phi \hat{n}}e^{i\theta \hat{n}}\\
    &= e^{i\phi k} \E{k}{\theta + \phi}.
\end{aligned}
\end{align}
These two expression can be combined concisely in the form $e^{i\phi \hat{n}} \E{k}{\theta} = e^{i\phi k \Theta(k)}\E{k}{\theta + \phi}$, which is the desired result.

%============================================
\section{Optimal implementation of a discrete rotation gate} \label{sec:rotation_implementation}
%============================================

In this section we detail how to construct a gate that acts as a $Z$-rotation of the angle $\phi_\ell = \pi/2^\ell$ in our codespace. Note that such a gate acting on a qubit has the matrix form
\begin{equation}
    \hat{R}_Z(\phi_\ell) = \left( \begin{array}{cc} 1 & 0 \\ 0 & e^{i\phi_\ell} \end{array} \right),
\end{equation}
so for our codespace we need a gate that does nothing to the bosonic Hilbert states $\ket{2kN}$ and applies a phase of $e^{i\phi_\ell}$ to the states $\ket{(2k + 1)N}$. This may be accomplished by letting $\hat{R}_N'(\phi_\ell) = \exp[if_\ell(\hat{n})]$, and our task is to find a suitable family of functions $f_\ell(n)$.

Explicitly, for a given $\ell$ we seek a function $f_\ell$ such that $f_\ell(kN)=2\pi m_k$ for $k$ even, and $f_\ell(kN)=2\pi m_k + \pi/2^\ell$ for $k$ odd, where $m_k$ can be any integer. It is easily checked that this will satisfy the conditions stated above. Rearranging arithmetically, we arrive at the following criteria for $f_\ell$:
\begin{equation} \label{eq:rotation_f_criteria}
    f_\ell'(k) = \left\{ \begin{array}{cc} 0 & k \text{ even} \\ 1 & k \text{ odd} \end{array} \right\} \mod 2^{\ell+1}
\end{equation}
where $f_\ell'(k) \equiv \frac{1}{\phi_\ell} f_\ell(kN)$ and $\phi_\ell = \pi/2^\ell$.

Modulo $2^{\ell+1}$, $f_\ell'(k)$ produces the sequence $S_k^{(0)} = f_\ell'(k) = 0,1,0,1,\dots$. Taking the difference of terms in this sequence produces the new sequence $S_k^{(1)} = S_{k+1}^{(0)} - S_k^{(0)} = (-1)^k$. Doing so again produces the sequence $S_k^{(2)} = S_{k+1}^{(1)} - S_k^{(1)} = -2(-1)^k$. We may continue this process all the way down to $S_k^{(\ell+1)} = (-1)^k (-2)^\ell$. Now we note that $2^\ell = -(2^\ell) \mod 2^{\ell+1}$, so $S_k^{(\ell+1)} = \text{const.} \mod 2^{\ell+1}$, and knowing that the $n^{\rm th}$ consecutive difference of a polynomial of degree $n$ is constant, we conclude that the criteria in \cref{eq:rotation_f_criteria} are satisfied by a polynomial of degree $\ell+1$. Since $S_k^{(n)} \ne \text{const.} \mod 2^{\ell +1}$ for $n \le \ell$, this must be the lowest order polynomial that satisfies the criteria.

Reversing the process above, we see that
\begin{align}
\begin{aligned}
    S_k^{(\ell)} &= S_0^{(\ell)} + \sum_{k'=0}^{k-1} S_{k'}^{(\ell+1)} &\mod 2^{\ell+1} \\
    &= S_0^{(\ell)} + k S_0^{(\ell+1)} \\
    &= {k \choose 0} S_0^{(\ell)} + {k \choose 1} S_0^{(\ell+1)} &\mod 2^{\ell+1}.
\end{aligned}
\end{align}
\begin{widetext}
Similarly,
\begin{align}
\begin{aligned}
    S_k^{(\ell-1)} &= S_0^{(\ell-1)} + \sum_{k'=0}^{k-1} S_{k'}^{(\ell)} &\mod 2^{\ell+1} \\
    &= S_0^{(\ell-1)} + \sum_{k'=0}^{k-1} \left( {k' \choose 0} S_0^{(\ell)} + {k' \choose 1} S_0^{(\ell+1)} \right) &\mod 2^{\ell+1} \\
    &= {k \choose 0} S_0^{(\ell-1)} + {k \choose 1} S_0^{(\ell)} + {k \choose 2} S_0^{(\ell+1)} &\mod 2^{\ell+1}.
\end{aligned}
\end{align}
Iterating this all the way back to the top yields
\begin{equation} \label{eq:f_series}
    S_k^{(0)} = \sum_{i=0}^{\ell+1} {k \choose i} S_0^{(i)} = \sum_{i=1}^{\ell+1} (-2)^{i-1} {k \choose i} \mod 2^{\ell+1},
\end{equation}
where the right-hand side is a polynomial in $k$ of degree $\ell+1$ which is a satisfying assignment to the function $f_\ell'(k)$.

To convert this back to the function $f_\ell(n) = \frac{\pi}{2^\ell}f_\ell'(n/N)$, we decompose ${k \choose i} = \frac{k!}{i!(k-i)!}$ and use $\Gamma$ as the analytic continuation of the factorial function so that we may write \eqref{eq:f_series} as a general function of the noninteger $n/N$. Returning to the original problem, we conclude that by choosing
\begin{equation}
    f_\ell(n) = \frac{\pi}{2^\ell}f_\ell'(n/N) = \pi \sum_{i=1}^{\ell+1} (-2)^{i-\ell-1} \frac{\Gamma \left( \frac{n}{N}+1 \right) }{(i!) \Gamma \left( \frac{n}{N}-i+1 \right)},
\end{equation}
the gate
\begin{equation}
    \hat{R}_N'(\phi_\ell) = \exp\left [ i f_\ell(\hat{n}) \right ]
\end{equation}
implements a logical $Z$-rotation of angle $\phi_\ell = \pi/2^\ell$ with the lowest order dependence on the operator $\hat{n}$. It should be noted that while we have found a polynomial of lowest degree that satisfies the criteria, the solution is not unique. In particular, the polynomial in the implementation of $\TN'$ $\left( \frac{1}{2}n^3 + \frac{1}{4}n^2 - \frac{1}{2}n \right)$ is not identical to $f_2(n) = \frac{1}{6}n^3 - \frac{3}{4}n^2 + \frac{5}{6}n$ but its degree is the same.
\end{widetext}

\end{document}